\documentclass[twocolumn,preprintnumbers,amssymb,pre,superscriptaddress]{revtex4}

\usepackage{graphicx}
\usepackage{dcolumn}
\usepackage{bm}
\usepackage{amstext}

\newcommand{\ehat}{\hat{\mathbf e}}

\begin{document}

\title{Model for Polymorphic Transitions in Bacterial Flagella}

\author{Srikanth V. Srigiriraju and Thomas R. Powers\\
\textit{Division of Engineering, Box D, Brown University,
Providence, RI 02912}} 

\date{16 August 2005}

\begin{abstract}

Many bacteria use rotating helical flagellar filaments to swim.
The filaments undergo polymorphic transformations in which the
helical pitch and radius change abruptly. These transformations
arise in response to mechanical loading, changes in solution
temperature and ionic strength, and point substitutions in the
amino acid sequence of the protein subunits that make up the
filament.  To explain polymorphism, we propose a new
coarse-grained continuum rod theory based on the quaternary
structure of the filament. The model has two molecular switches.
The first is a double-well potential for the extension of a
protofilament, which is one of the eleven almost longitudinal
columns of subunits. Curved filament shapes occur in the model
when there is a mismatch strain, i.e. when inter-subunit bonds in
the inner core of the filament prefer a subunit spacing which is
intermediate between the two spacings favored by the double-well
potential. The second switch is a double-well potential for twist,
due to lateral interactions between neighboring protofilaments.
Cooperative interactions between neighboring subunits within a
protofilament are necessary to ensure the uniqueness of helical
ground states. We calculate a phase diagram for filament shapes
and the response of a filament to external moment and force.

\end{abstract}



\maketitle

\section{Introduction}

Allosteric proteins switch between different conformations in
response to the binding of ligands. Allostery underlies a wide
range of cellular processes, such as the cooperative binding of
oxygen to hemoglobin~\cite{eaton_etal1999}, the negative feedback
control of the DNA-binding \textit{trp} repressor in
\textit{Escherichia coli}~\cite{lawsonsigler1988}, and the
regulation of the catalyst aspartate transcarbamoylase, also in
\textit{E. coli}~\cite{krausevolzlipscomb1985}. All these examples
involve conformational changes of a small molecule in response to
the the binding of a ligand. Bray and Duke have recently argued
that the concept of allostery should be extended from small
molecules to large assemblies of proteins, in which conformational
change propagates through a one- or two-dimensional
lattice~\cite{brayduke2004}. In this picture of ``conformational
spread," the state of an individual protein subunit of a large
assembly is stabilized either by ligand binding \textit{or} by the
conformational state of a neighboring subunit.

In this article we develop a theory based on conformational spread
for polymorphic transformations of the flagellar filaments of
bacteria such as \textit{E. coli}. These filaments are
hollow tubes built from identical protein subunits known as
flagellin~\cite{macnab1996}. To explain why the tubes are usually
helical, instead of straight cylinders, Asakura supposed that each
flagellin subunit can be in one of two distinct conformations,
with one slightly longer than the other~\cite{asakura1970}. Since
the subunits may be grouped into eleven protofilaments which
slowly wind around each other to form the filament, a helical
filament shape arises when some of the protofilaments consist of
only long subunits, and the rest consist of only short subunits.
For example, for physiological solution conditions and in the
absence of external forces and moments, the filament of \textit{E.
coli} is left-handed with a pitch of $2.5$\,$\mu$m and a diameter
of $0.4$\,$\mu$m~\cite{turner_ryu_berg2000}. Indirect evidence of
the spread of conformational change comes from the response of a
helix to hydrodynamic torque, which can trigger a polymorphic
transition from a left-handed normal state to a right-handed
state~\cite{macnab_ornston1977}. For example, Hotani has detached
filaments from \textit{Salmonella typhimurium} cells and used hydrodynamic
torque to trigger polymorphic transformations~\cite{hotani1982}. A
rotating helix in a viscous fluid generates propulsive thrust;
conversely, flow directed along the axis of a helix not only
stretches the helix, but also tends to untwist the helix,
independent of its handedness. By flowing solution past filaments
stuck at one end to a microscope slide, Hotani observed different
helical states propagating from the stuck end to the free
end~\cite{hotani1982}.

Polymorphic transitions caused by hydrodynamic torque also occur
in free-swimming bacteria~\cite{macnab_ornston1977}. There are
usually five to six flagella per cell. When the flagellar motors turn
counterclockwise (as viewed from outside the cell), the
left-handed helices wrap into a bundle which propels the cell
forward~\cite{berg2004,kim_etal2003}. Reversal of the rotation
direction of a motor leads to a sequence of events in which the
corresponding filament unwinds from the bundle and transforms to a
right-handed state~\cite{macnab_ornston1977}, before eventually
wrapping up into the bundle once the motor returns to
counterclockwise rotation. Fluorescent labelling of the filaments
has recently revealed the precise dynamics of this
process~\cite{turner_ryu_berg2000}. The typical sequence begins
with a transformation from the left-handed normal state to a
right-handed ``semi-coiled" state with half the helical pitch of
the normal state. This transformation changes the swimming
direction of the cell. Then there is an additional transformation
to a right-handed state called ``curly-1," which has a similar
pitch but roughly half the radius of the semi-coiled state. When
the motor reverses back to counterclockwise rotation, the filament
transforms directly to the normal state.

Discontinuous transitions among these states can also occur due to
changes in pH~\cite{kamiyaasakura1976}, salt
concentration~\cite{kamiyaasakura1976}, or
temperature~\cite{hasegawa_etal1982}. Straight
states~\cite{asakura1970} and curved planar (``coiled")
states~\cite{kamiyaasakura1976} have also been observed.

In this article we describe a continuum model for flagellar
filaments. We begin in section~\ref{prevwork} with a review of
experiments that reveal the structure and mechanical response of
the flagellar filament, and then discuss previous theoretical
models for polymorphism.  We introduce our model in
section~\ref{ourenergy}. The essential features of our model are
two inequivalent subunit conformations, a mismatch of the
preferred spacing of protein domains in the inner and outer core
of the filament, and cooperative interactions between neighboring
subunits. All three of these features are suggested by the
experiments, and are required for our model to yield a unique
helical ground state for given parameters. In
section~\ref{simpmod} we study a simplified version of our model,
introduced in~\cite{SrigirirajuPowers2005}, which neglects
extensibility and twist-stretch interactions. Using this model, we
calculate the filament ground states as a function of material
parameters. Then we turn to the response of the filament to
external moment. Some of these results were described
in~\cite{SrigirirajuPowers2005}. We conclude this section with new
calculations of the response to external force. In
section~\ref{fullmod}, we present the phase diagram and response
to external moments and forces for the full model, in which the
simplifying assumptions of section~\ref{simpmod} are removed. We
discuss our results in section~\ref{discusssec}.

\begin{figure}
\includegraphics[height=1.4in]{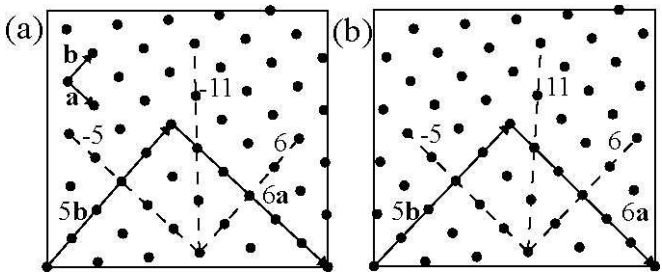}
\caption{The lattices of (a) L-type and (b) R-type filaments, at
radius $a=4.5$\,nm and using data from~\cite{yamashita_etal1998}.
The vertical edges are identified. Note the 11-, 6-, and 5-start
helices. The protofilaments are the 11-start helices.}
\label{lattices}
\end{figure}

\begin{figure}
\includegraphics[height=2in]{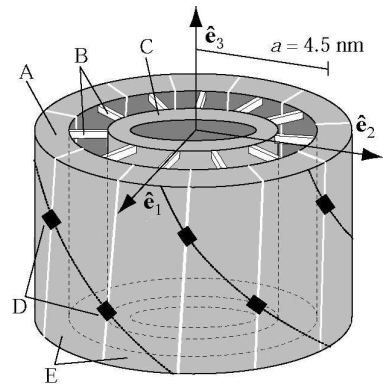}
\caption{Cartoon of a short segment of the R-type flagellar
filament, showing the outer core (A), connecting spokes (B), inner
core (C), and lateral bonds (D) between neighboring protofilaments
(E). The position of the spokes is not accurate, and is only meant
to indicate that the inner and outer cores are connected. The
outer domains are not shown. The lateral bonds lie along the five
5-start helices. } \label{crossfig}
\end{figure}

\section{Summary of previous work.}\label{prevwork}

Before the detailed structures of virus capsids were measured, it
was argued that self-assembly of identical subunits would lead to
highly symmetric forms, with all subunits in equivalent bonding
environments~\cite{crick_watson1956}. For example, the protective
coat of tobacco mosaic virus consists of identical protein
subunits which form a straight
cylinder~\cite{fraenkel-conrat_williams1955}. In flagella, the
flagellin subunits also form a two-dimensional crystal on the
surface of a cylinder, which is about 20\,nm in diameter. As
mentioned already, the subunits can be grouped into eleven
protofilaments that gradually wind around each other to form the
filament. Since the flagellum usually has a helical shape, exact
equivalence cannot apply. However, since the filament is thin,
small departures from exact equivalence can lead to helices of the
observed dimensions~\cite{asakura1970}.

\subsection{Experiments}
\label{expts}

A consequence of Asakura's picture~\cite{asakura1970} of two stable states for
protofilaments is that two distinct straight filaments are
possible, one in which all protofilaments are short, and one in
which all are long. These states, known as R-type and L-type, have
been observed~\cite{hyman_trachtenberg1991}. The subunit period
along a protofilament is $0.8$\,\AA~shorter in R-type than in
L-type. The protofilaments make a positive angle with the
longitudinal direction in R-type, leading to right-handed twist,
whereas the protofilaments in L-type have a left-handed twist.
Figures~\ref{lattices} (a) and (b) show the subunit lattices in L-
and R-type at a radius of 4.5\,nm, obtained from x-ray scattering
from fibers of filaments~\cite{yamashita_etal1998}. The primitive
lattice vectors $\mathbf{a}$ and $\mathbf{b}$ in the two figures
are slightly different, but in both cases
$6\mathbf{a}+5\mathbf{b}=C\hat\mathbf{x}$ where $C$ is the
circumference of the filament at 4.5\,nm, and
$C\hat\mathbf{x}=\mathbf{0}$ since the lattices lie on the surface
of a cylinder. The vector constraint on the two lattice vectors
yields two degrees of freedom, which may be taken to be the
subunit spacing along the 11-start direction and the angle the
11-start direction makes with a longitude. Equivalently, we may
replace the angle degree of freedom with the twist of the
filament.

Electron microscopy has shown that the filament cross-section
consists of outer domains surrounding a core with inner and outer
parts (Fig.~\ref{crossfig})~\cite{mimori_etal1995}. The inner
and outer cores are connected by radial spokes.  The outer
domains do not seem to affect the mechanical properties of the
filament; for example, mutations in the outer domain do not affect
polymorphism~\cite{yoshioka_aizawa_yamaguchi1995}. Both the outer
core and the inner core are crucial for polymorphism. Mutations in
the outer core can change the ground state of the filament from
normal to curly, semi-coiled, or straight~\cite{kanto_etal1991}.
Without the inner core, there is no polymorphic behavior: the
filaments are straight, and the subunit lattice at radius
$4.5$\,nm is of Lt-type, with left-handed twist and subunits in
the short state~\cite{mimori-kiyosue_etal1996}. (But note that
subunits without the inner region can polymerize to form helical
filaments if there is a wild-type helical
seed~\cite{vonderviszt_aizawa_namba1991}).

Since flagellin tends to self-assemble into filaments, it is
difficult to crystallize. By clipping off some of the residues at
either end of a flagellin chain, Samatey et al. crystallized the
F41 fragment of flagellin from the R-type filament of \textit{S.
typhimurium}, allowing $2.0$\,\AA~resolution of the subunit
structure by x-ray scattering~\cite{samatey_etal2001}. The
crystals consist of an anti-parallel arrangement of R-type
protofilaments. The lateral interactions normally present in an 
intact filament are different in the crystal, but the 
protofilament is still stable. Therefore,
the authors of~\cite{samatey_etal2001} suggested that the
protofilament is an ``independent, cooperatively switching unit."
The structure of the intact R-type filament was revealed at the
$4$\,\AA~resolution level by electron
cryomicroscopy~\cite{yonekura_etal2003}. This work showed that
each subunit interacts with its neighbors in the outer core along
the 5-, 11-, and 16-start directions, but not along the 6-start
direction. In the inner core, each subunit contacts its neighbors
along the 5-, 6-, and 11-start directions.

\subsection{Models}
The first quantitative theory of polymorphism in flagella is due
to Calladine, who modeled an element of a filament as two rigid
discs connected by eleven linear springs~\cite{calladine1975,
calladine1976,calladine1978}. Following Asakura, he supposed that
the springs could have one of two rest lengths, with the same
spring constant in either case. If all the springs have the same
rest length, then a stack of elements would be straight. If some
are short and the rest are long, then in equilibrium there will be
a variation of spring length, causing the discs in an element to
incline relative to each other. A stack of these elements will
form a curved filament in equilibrium, with the curvature varying
sinusoidally with the number of long springs~\cite{calladine1976}.

Calladine further supposed that the straight state with long
springs has the twist of the L-state, the straight state with
short springs has the twist of the R-state, and a filament with a
mix of short and long springs has a twist that linearly
interpolates between the two extremes. These assumptions lead to
ten discrete helical states with curvature varying sinusoidally
with twist. A key assumption of this model is that the torsion of
a helical state is equal to the twist of the elements. Calladine's
model was studied further by Hasegawa et al.~\cite{hasegawa_etal1998}, who used
more accurate data to determine the twist parameters of the model,
and who also calculated the range of deviations in subunit spacing
in single filament of a given helical shape. The predictions of
Calladine's model for the possible ground states of flagellar
filaments agree reasonably well with measurements of the filament
shape~\cite{yamashita_etal1998}.

Goldstein et al. used a different approach to model flagellar
filaments~\cite{goldstein_etal2000}. They presented a continuum
theory for an elastic rod with a preferred direction of curvature,
a preferred magnitude of curvature, and a double-well potential
for twist. Although the preferred direction of curvature is hard
to reconcile with the fact that the flagellin subunits are
identical, this model is probably the simplest continuum model for
bistable helices. Unlike the model
of~\cite{calladine1975,calladine1976,calladine1978}, it provides a
framework for calculating the deformation of the filament in
response to external forces and moments, such as hydrodynamic
loading~\cite{CoombsHuberKesslerGoldstein2002}. This continuum
model, along with resistive force theory for hydrodynamic drag,
leads to a prediction for the velocity of the front propagating
along a helix during a transition from one polymorphic form to
another~\cite{CoombsHuberKesslerGoldstein2002}.

In this article we develop a new model which has elements similar
to both the spring model and the continuum bistable helix
model~\cite{SrigirirajuPowers2005}. Like the model of Goldstein
and collaborators, our model is a continuum rod theory that
predicts the deformation in response to load. Like Calladine's
model, our model is based on the two conformations of the protein
subunits. However, our model introduces important new features.
For example, it does not single out a preferred direction of
curvature, but instead assigns every protofilament the same
double-well potential for stretch. The model incorporates
pre-stress by assuming there is an elastic mismatch between the
inner and outer cores of the filament; this elastic mismatch is
necessary to make the absolute minimum of the elastic energy
helical instead of straight. Finally, we explicitly treat the
cooperative interactions between neighboring subunits. These
cooperative interactions are required to make the ground state
unique.

\section{Filament elastic energy}\label{ourenergy}
The goal of this article is to explain why flagellar filaments are
helical and why they undergo polymorphic transformations. Since
the characteristic length scale of the helical polymorphs is a
micron, much larger than the 5\,nm scale of the flagellin
subunits, it is natural to formulate a coarse-grained or continuum
theory for polymorphism. We model the filament as two concentric
elastic cylinders (Fig.~\ref{crossfig}). The inner cylinder
corresponds to the inner core of the filament, and is treated as
an elastic rod with a resistance to stretching. The outer cylinder
corresponds to the outer core of the filament, and consists of
eleven strands which gently wind around the inner core
(Fig.~\ref{crossfig}). Each strand corresponds to a protofilament,
although strictly speaking, a protofilament in a flagellar
filament includes material from both the inner and outer cores. We
divide the outer core into protofilaments but treat the inner core
as a single elastic element since only the outer core is
implicated in switching between polymorphic
forms~\cite{kanto_etal1991}.

Our model treats the eleven strands and the inner core as
continua. To define the strains of the strands and the inner core,
we choose a reference configuration and label the points along the
inner core with arclength $S$ in the reference configuration. The
strain of the inner core is $\epsilon=\mathrm{d}s/\mathrm{d}S-1$,
where $s$ is the actual arclength of the inner core. Note that the
strain need not vanish when the inner core is in the reference
configuration. Since the difference in subunit spacing of L- and
R-type is about 2\% of the subunit size, we assume $\epsilon\ll1$.
To define the strain $\epsilon_i$ of the $i$th strand, where
$i=1,2,\ldots,11$, consider the cross-section of the filament at
$S$. A point on a strand that intersects this cross-section is
assigned the label $S$. Thus, if $\mathrm{d}\ell_i$ is the element
of length of the $i$th strand, then
$\epsilon_i=\mathrm{d}\ell_i/\mathrm{d}S-1$.

Our model treats the inner core as a rod with a quadratic strain energy     
per unit length,
\begin{equation}
U_0=\frac{1}{2}k_\mathrm{s}(\epsilon-\epsilon_0)^2, \label{U0eqn}
\end{equation}
where $k_\mathrm{s}$ is an elastic constant and $\epsilon_0$ is       
the strain of the inner core in its unstressed state. The total
stretching energy of the inner core is $\int U_0(S)\;\mathrm{d}S$.
But since $\mathrm{d}S=\mathrm{d}s/(1+\epsilon)\approx\mathrm{d}s$
to an excellent approximation, we may approximate all energies as
integrals over $s$ rather than $S$. The distinction between $s$
and $S$ must be maintained only when computing strain $\epsilon$.

\begin{figure}
\includegraphics[height=1.8in]{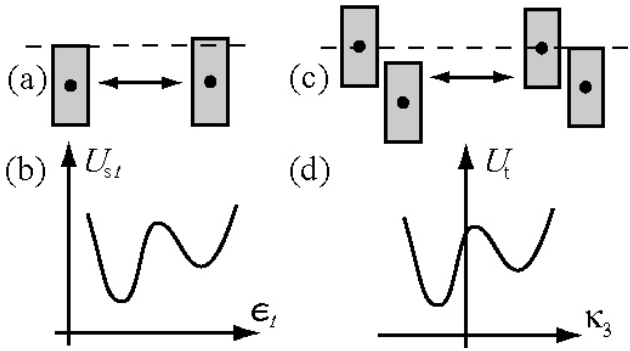}
\caption{(a) The two preferred conformations of the subunit. (b)
Sketch of double-well potential for stretching a protofilament,
with minima corresponding to the two preferred states of (a). (c)
The two preferred relative positions of neighboring subunits. (d)
Sketch of double-well potential for twist $\kappa_3$. }
\label{potentialsfig}
\end{figure}

Motivated by the discussion of two preferred subunit
conformations, we introduce a ``switch" for protofilament
extension by defining the stretching energy per unit length of a
strand as a double-well potential,
\begin{equation}
U_{\mathrm{s}i}=u\left[(\epsilon_i^2-\epsilon_\mathrm{p}^2)^2/4
-u_1\epsilon_\mathrm{p}^3\epsilon_i\right], \label{Usieqn}
\end{equation}
where $u$ is an elastic constant, $\epsilon_\mathrm{p}$ determines
the positions of the local minima of $U_{\mathrm{s}i}$ that
correspond to the two conformations, and the dimensionless
parameter $u_1$ determines the degree of asymmetry of
$U_{\mathrm{s}i}$ [Fig.~\ref{potentialsfig}(a), (b)]. A term
proportional to $\epsilon_i^3$ is disregarded for simplicity.

Since radial spokes connect the inner and outer cores, the strains
$\epsilon$ and $\epsilon_i$ are not independent. For example, when
the filament is straight, $\epsilon=\epsilon_i$. Mismatch strain
arises because the preferred strain of the inner core need not
agree with the either of the strains corresponding to the minima
of $U_{\mathrm{s}i}$. We will vary the mismatch strain by changing
$\epsilon_0$ and keeping $U_{\mathrm{s}i}$ fixed. Mismatch strain
is a natural assumption since during self-assembly, the subunits
must denature as they travel up the narrow channel at the core of
a filament before inserting into their final destination at the
distal end~\cite{berg2004}.

When the filament is curved, the extensions $\epsilon_i$ depend on
$i$. To determine this dependence, we follow Calladine and make
the simplest assumption by supposing that planar cross-sections of
the filament remain planar under bending and
twisting~\cite{calladine1976}. We further suppose that the rod is
unshearable. Thus the position of a material point on the $i$th
strand at radius $a$ is
\begin{equation}
\mathbf{r}_i=\mathbf{r}_\mathrm{c}+ a\cos\beta_i\hat\mathbf{e}_1 +
a\sin\beta_i\hat\mathbf{e}_2,\label{rieqn}
\end{equation}
where $\mathbf{r}_\mathrm{c}$ is the position of the centerline of
the inner core, $\beta_i=2\pi(i-1)/11$, and $\hat\mathbf{e}_1$ and           
$\hat\mathbf{e}_2$ are orthonormal material frame vectors which
are perpendicular to the tangent vector
$\hat\mathbf{e}_3=\mathrm{d}\mathbf{r}_c/\mathrm{d}s$
(Fig.~\ref{crossfig}). Without loss of generality we have chosen
$\ehat_1$ to point to the protofilament with $i=1$.

In this article we choose $a=4.5$\,nm, the
radius at which the lattice structures of Fig.~\ref{lattices} are
reported. Since
$\{\hat\mathbf{e}_1,\hat\mathbf{e}_2,\hat\mathbf{e}_3\}$ form an
orthonormal frame,
\begin{equation}
\frac{\mathrm{d}\hat\mathbf{e}_\mu}{\mathrm{d}s}=\bm{\kappa}
\times\hat\mathbf{e}_\mu,\label{onframeeqn}
\end{equation}
where $\bm{\kappa}=\kappa_\mu\hat\mathbf{e}_\mu$, $\kappa_1$ and
$\kappa_2$ are the components of the curvature vector
$\bm{\kappa}_\perp=\mathrm{d}\hat\mathbf{e}_3/\mathrm{d}s =
\kappa_2\hat\mathbf{e}_1- \kappa_1\hat\mathbf{e}_2$, and
$\kappa_3$ is the twist. The components $\kappa_1$ and $\kappa_2$ are the 
rates that the orthonormal frame rotates about $\hat{\mathbf e}_1$ and 
$\hat{\mathbf e}_2$, respectively. Likewise, the twist $\kappa_3$ is the 
rate that the orthonormal frame rotates about $\hat{\mathbf e}_3$. 
Note again that since $\epsilon\ll1$,
whether we use $\mathrm{d}/\mathrm{d}s$ or
$\mathrm{d}/\mathrm{d}S$ to define $\bm{\kappa}$ is immaterial.
The magnitudes of the components of $\bm{\kappa}$ are much smaller
than $1/a$: $|\kappa_\mu| a\ll1$. Thus, the relations
(\ref{rieqn}) and (\ref{onframeeqn}) together with
$\mathrm{d}\ell_i=(\mathrm{d}\mathbf{r}_i/\mathrm{d}S
\cdot\mathrm{d}\mathbf{r}_i/\mathrm{d}S)^{1/2}\mathrm{d}S$ imply
\begin{equation}
\epsilon_i=\frac{\mathrm{d}\ell_i-\mathrm{d}S}{\mathrm{d}S}
\approx\epsilon + a \kappa_1\sin\beta_i-a\kappa_2 \cos\beta_i.
\label{epieqn}
\end{equation}
Although the strain energy of Eq.~(\ref{Usieqn}) involves terms of
up to fourth order in $\epsilon_i$, the form of the coefficients
of the quartic potential justifies approximating $\epsilon_i$ to
first order in $\epsilon$ and $\kappa_\mu a$.

Note that the twist $\kappa_3$ does not appear to first order
since the protofilaments are almost longitudinal. For example, at
radius 4.5\,nm, the angle between a protofilament and the
longitude is $-1.45^\circ=-0.0253$\,rad in L-type and
$3.46^\circ=0.0605$\,rad in R-type~\cite{yamashita_etal1998}.

The stretching energy of all the strands is
$U_\mathrm{s}=\Sigma_{i=1}^{11}U_{\mathrm{s}i}$, or
\begin{equation}
\frac{U_\mathrm{s}}{11}=\frac{u}{4}\left[(\epsilon^2-\epsilon_\mathrm{p}^2)^2
+(3\epsilon^2-\epsilon_\mathrm{p}^2)a^2\kappa^2 +
\frac{3}{8}a^4\kappa^4\right]-u_1\epsilon_\mathrm{p}^3\epsilon
\end{equation}
where $\kappa=|\bm{\kappa}_\perp|>0$ is the curvature of the
filament. Although $U_{\mathrm{s}i}$ has a term linear in
$\epsilon_i$, there are only even powers of $\kappa$ in
$U_\mathrm{s}$. Thus, the energy $U_\mathrm{s}$ depends on the
magnitude of the curvature. There is no natural (or spontaneous)
\textit{direction} of curvature since all subunits are identical.
However, the subunits may be in different states. For example, if
$3\epsilon^2-\epsilon_\mathrm{p}^2<0$, then there will be a
natural \textit{magnitude} of curvature, and a distribution of
subunit states around the circumference of the filament. In this
case, the direction of the curvature vector at a point $s$ may
point in any direction in the plane of the cross-section. Note
also that $U_\mathrm{s}$ shows no trace of the original
eleven-fold symmetry of the sum of $U_{\mathrm{s}i}$ over the
eleven protofilaments; the resistance to bending is isotropic for
a rod with a cross-section with a symmetry greater than
two-fold~\cite{landau_lifshitz_elas}. This isotropy rules out
couplings between twist and curvature like those that arise in the
mechanics of DNA~\cite{marko_siggia1994}.

\begin{figure}
\includegraphics[height=1.4in]{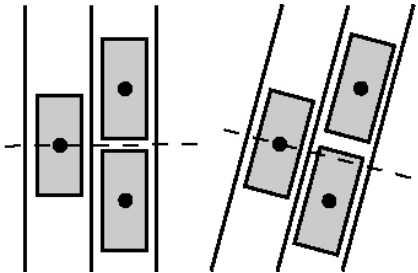}
\caption{Illustration of how twist causes neighboring subunits to
slide past each other. The subunits rotate rigidly but the lattice
of centers of mass shears. The angle of shear has been exaggerated
for illustration. } \label{sheartwistfig}
\end{figure}

In our model, lateral interactions between neighboring subunits on
different protofilaments lead to a resistance to twist. We follow
the suggestion of Namba and
Vonderviszt~\cite{namba_vonderviszt1997} and suppose that
neighboring protofilaments have two preferred relative
displacements along the local protofilament direction
[Fig.~\ref{potentialsfig}(c), (d)]. These interactions are at the
quaternary level; they do not require a conformational change of
the subunit. As Fig.~\ref{sheartwistfig} illustrates, twisting a
filament causes neighboring protofilaments to slide past each
other.  Therefore, two preferred relative displacements are
equivalent to two preferred twists. Thus, we introduce a second
switch, a double-well potential for twist:
\begin{equation}
U_\mathrm{t}=v\left[\frac{1}{4}a^4(\kappa_3^2-\Omega_\mathrm{p}^2)^2-\bar{v}_1
a^4 \Omega_\mathrm{p}^3 \kappa_3\right],\label{Ut}
\end{equation}
where $v$ is an elastic constant, $\Omega_\mathrm{p}$ determines
the two preferred twists, and $\bar{v}_1$ makes the potential
asymmetric. As for our potential for extension,
Eq.~(\ref{Usieqn}), a cubic term is disregarded for simplicity.

In addition to the switches for extension and twist, the filament
structure suggests a coupling between twisting and stretching. As
mentioned earlier, the electron cryomicroscopy studies of the
intact R-type filament show that each subunit has contacts with
its neighbors along the 5-, 11-, and 16-start directions in the
outer core. Our model effectively replaces all the inner core
lattice contacts with the inner tube, and those along the 11-start
direction of the outer tube by a double-well potential. Since the
16-start direction is almost longitudinal, like the 11-start
direction, we ignore it for simplicity. The 5-start helices make
an angle of approximately $-45^\circ$ with the longitude in both
the straight filaments. This contact will lead to twist
resistance, as well as a coupling between twist and stretch. To
understand the sign of this coupling, consider a straight R-type
filament in the limit where the bonds along the 5-start direction
are rigid and unable to stretch. Pulling on the filament will
cause the left-handed 5-start helices to untwist a little, leading
to an increase in $\kappa_3$, the twist of the 11-start helices.
Likewise, an applied moment which tends to unwind the 5-start
helices will cause the filament to lengthen. Thus, the twist
stretch coupling is of the form
\begin{equation}
U_5=-k_5 a \epsilon\kappa_3, \label{U5}
\end{equation}
where $k_5$ is a positive elastic constant.

The 5- and 6-start contacts in the inner core will also contribute
to the twist-stretch coupling. Without knowing the relative
strength of these bonds, we cannot predict the sign of the
contribution to the twist-stretch coupling, since the 5-start
helices are left-handed and the 6-start helices are right-handed.
However, since the outer core bonds lie at a radius roughly double
that of the inner core bonds, and bending and twisting moduli in
rod theory scale as the third power of radius for a hollow
rod~\cite{landau_lifshitz_elas}, we assume that the contribution
to the elastic energy of the inner core bonds is dominated by that
of the outer core bonds. Note also that the 5-start bonds will
contribute to the bending and twisting resistance of the filament;
these effects are already accounted for in the bending terms of
$U_\mathrm{s}$ and the twist potential $U_\mathrm{t}$. Also, since
the high-resolution structure of the intact L-type filament is
currently unavailable, we make the simplest assumption and suppose
there are contacts along the 5-start direction but not the 6-start
direction in L-type as well as R-type. Finally, we note that there
could be a twist-stretch coupling arising from the elasticity of
the subunit itself. For example, pulling on a subunit along the
11-start direction could cause it to shear as well as stretch,
causing the whole filament to twist in a right or left-handed
sense. We know of no evidence for or against this possibility, but
we must acknowledge that our choice of the sign of $k_5$ is
provisional.

\begin{figure}
\includegraphics[height=2.0in]{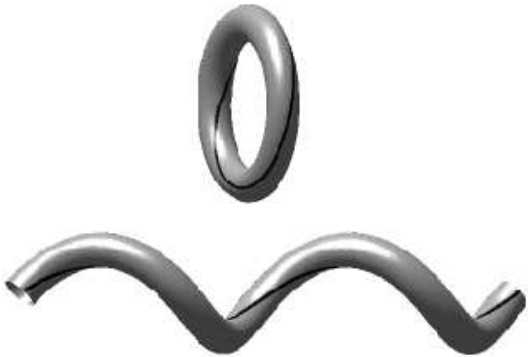}
\caption{Two rods with the same curvature and the same twist. The
black line traces the path of the head of the vector
$\hat\mathbf{e}_1$. The circular rod is twisted through one turn to
make $\kappa_3=1/R_\mathrm{c}$, where $R_\mathrm{c}=1/\kappa$ is
the radius of the circle.} \label{circlehelix}
\end{figure}

The terms of the elastic energy introduced above ($U_\mathrm{s}$,
$U_\mathrm{t}$, and $U_5$) determine the extension $\epsilon$, the
curvature $\kappa$, and the twist $\kappa_3$. However, these
variables do not completely determine the shape of the filament:
two rods may have the same curvature and the same twist, but
different shapes. For example, Fig.~\ref{circlehelix} shows
circular and helical filaments that have the same curvature and
the same twist, and hence the same value of
$U_\mathrm{s}+U_\mathrm{t}+U_5$ (assuming $\epsilon$ is the same
for both filaments). To see what kind of term is required to
remove this degeneracy, consider the special case of a rod bent
into a circle. In Fig.~\ref{circlehelix}, the curvature vector of
the circle lies in the plane of the circle, pointing toward the
center. Since the twist is nonzero, $\kappa_3\neq0$, the components of the
curvature vector in the material frame depend on arclength,
$\mathrm{d}\kappa_{1,2}/\mathrm{d}s\neq0$. Therefore, by
Eq.~(\ref{epieqn}), the extension $\epsilon_i$ of the $i$th
protofilament also depends on arclength. Contrast this example
with the case of a circular rod with no twist, $\kappa_3=0$. In this case,
$\kappa_1=0$ and $\kappa_2=1/R$, and the extension $\epsilon_i$ of
each protofilament is uniform. The shortest protofilament lies a
circle of radius $R-a$, and the longest protofilaments each lie on
a circle with radius slightly less than $R+a$ (since there is an
odd number of protofilaments). The helix in Fig.~\ref{circlehelix}
is analogous to the circle with no twist; the pitch and radius
have been carefully chosen to ensure that $\epsilon_i$ is uniform.

Motivated by these observations, we suppose there is an energy
penalty when neighboring subunits on the same protofilament are in
different states. This cooperative interaction is implicit in the
assumptions of Asakura~\cite{asakura1970} and
Calladine~\cite{calladine1975}, who supposed all subunits in a
given protofilament prefer to be in the same state. Also, as
mentioned in section~\ref{expts}, the integrity of the
protofilaments in crystallized R-type subunits led Samatey and
collaborators to suppose there are strong cooperative interactions
along the 11-start direction~\cite{samatey_etal2001}. In our
continuum model, the cooperative energy penalizes nonuniform
extension of each protofilament:
\begin{equation}
U_\mathrm{c}=\frac{wa}{2}\sum_{i=1}^{11}\left(\frac{\mathrm{d}\epsilon_i}{\mathrm{d}s}\right)^2,
\label{coop}
\end{equation}
where $w$ is an elastic constant with units of energy per unit length. 
The cooperative interaction Eq.~(\ref{coop}) stabilizes the state of a 
given subunit, depending on the state of the subunits neighbors, in accord 
with the concept of conformational spread~\cite{brayduke2004}. In terms of
the filament variables $\epsilon$, $\kappa_1$, and $\kappa_2$, the
cooperative interaction is
\begin{equation}
U_\mathrm{c}=\frac{11wa}{2}\left[\left(\frac{\mathrm{d}\epsilon}{\mathrm{d}s}\right)^2
+\frac{a^2}{2}\left(\frac{\mathrm{d}\kappa_1}{\mathrm{d}s}\right)^2
+
\frac{a^2}{2}\left(\frac{\mathrm{d}\kappa_2}{\mathrm{d}s}\right)^2\right].
\label{coopk1k2}
\end{equation}

\begin{figure}
\includegraphics[height=2.0in]{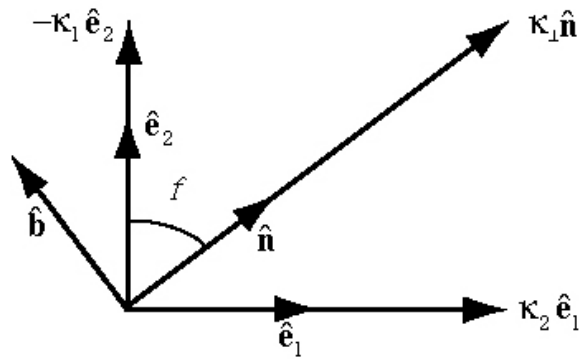}
\caption{The basis vectors of the material and Serret-Frenet
frames in the plane of the rod cross-section.} \label{fsmatfig}
\end{figure}

It is also convenient to write $U_\mathrm{c}$ in terms of the
Serret-Frenet basis~\cite{kamien2002}. Recall that the
Serret-Frenet basis of a space curve is an orthonormal frame
consisting of the tangent
$\hat\mathbf{t}=\hat\mathbf{e}_3=\mathrm{d}\mathbf{r}/\mathrm{d}s$,
normal $\hat\mathbf{n}$, and binormal $\hat\mathbf{b}$, where
\begin{equation}
\frac{\mathrm{d}}{\mathrm{d}
s}\left(\matrix{\hat\mathbf{n}\cr\hat\mathbf{b}\cr\hat\mathbf{t}}\right)=
\left(\matrix{0&\tau&-\kappa\cr-\tau&0&0\cr\kappa&0&0}\right)
\left(\matrix{\hat\mathbf{n}\cr\hat\mathbf{b}\cr\hat\mathbf{t}}\right),
\label{FS}
\end{equation}
and $\tau$ is the torsion. (Recall also that unlike $\tau$,
$\kappa_1$, $\kappa_2$, and $\kappa_3$, each of which may have
either sign, the curvature $\kappa$ is nonnegative.) Since
$\{\hat\mathbf{n},\hat\mathbf{b}\}$ and
$\{\hat\mathbf{e}_1,\hat\mathbf{e}_2\}$ span the same plane, they
are related by a rotation:
\begin{eqnarray}
\hat\mathbf{e}_1 &=&\sin f \hat\mathbf{n}-\cos f \hat\mathbf{b}\\
\hat\mathbf{e}_2 &=&\cos f \hat\mathbf{n}+\sin f \hat\mathbf{b}.
\end{eqnarray}
These relations, together with
$\bm{\kappa}_\perp=\mathrm{d}\hat\mathbf{e}_3/\mathrm{d}s$, imply
$\kappa_1=-\kappa\cos f$ and $\kappa_2=\kappa\sin f$ (see
Fig.~\ref{fsmatfig}). The relation between twist and torsion may
be found by writing $\kappa_3=-\hat
\mathbf{e}_1\cdot\mathrm{d}\hat\mathbf{e}_2/\mathrm{d}s$ in terms
of the Serret-Frenet basis:
\begin{equation}
\kappa_3=\mathrm{d}f/\mathrm{d}s+\tau. \label{kappa3fstau}
\end{equation}
Thus,
\begin{equation}
U_\mathrm{c}=\frac{11wa}{2}\left[\left(\frac{\mathrm{d}\epsilon}{\mathrm{d}s}\right)^2
+ \frac{a^2}{2}\left(\frac{\mathrm{d}\kappa}{\mathrm{d}s}\right)^2
+ \frac{a^2}{2}\left(\tau-\kappa_3\right)^2\right].
\label{coopktauf}
\end{equation}
The cooperative term $U_\mathrm{c}$ is at its absolute minimum
when the extension and curvature of the filament are uniform, and
when the torsion is equal to the twist. Thus, $U_\mathrm{c}$
removes the degeneracy in the filament ground states. When
$\tau=\kappa_3$ [as for the helix of Fig.~\ref{circlehelix}, with
radius $R=\kappa/(\kappa^2+\kappa_3^2)$ and pitch
$P=2\pi\kappa_3/(\kappa^2+\kappa_3^2)$], the angle $f$ is
constant, and therefore the Serret-Frenet frame does not rotate
relative to the material frame as $s$ changes. Without loss of
generality we may consider $f=\pi/2$, in which case the
Serret-Frenet frame and the material frame align for all $s$.
Therefore, the protofilament $i=1$ traces out the path
$\mathbf{r}(s)+a\mathbf{n}(s)$; this path is the ``shortest path"
on the surface of the helical filament, analogous to the path of
radius $R-a$ on the surface of a circular tube with centerline
radius $R$ and tube radius $a$.

To summarize, the total elastic energy is
\begin{equation}
E=\int(U_0+U_\mathrm{s}+U_\mathrm{t}+U_5+U_\mathrm{c})\mathrm{d}s.\label{E}
\end{equation}
The material parameters, such as $\epsilon_0$ and $\bar v_1$, for
example, are functions of environmental conditions such as pH and
temperature. These functions must be determined by a microscopic
theory and are therefore outside the scope of our coarse-grained
theory. Therefore, in the following sections we will determine the
filament shape as a function of the material parameters and
externally imposed moments and forces.

\section{Simple model: stiff central spring}\label{simpmod}


First we consider the simplified model
studied in~\cite{SrigirirajuPowers2005}, in which the stretching
stiffness of the inner core is large
($k_\mathrm{s}/u\rightarrow\infty$), the nonlinear stretching
potential is symmetric ($u_1=0$), and the twist-stretch coupling
is disregarded ($k_5=0$).  Since the central core is so stiff, the
extension of the filament is $\epsilon=\epsilon_0$, and the total
energy density (up to an unimportant additive constant) simplifies
to
\begin{eqnarray}
U&=&\frac{\tilde
u}{4}a^4\left(\kappa^2-\kappa_0^2\right)^2\nonumber\\
&+&v\left[\frac{a^4}{4}\left(\kappa_3^2-\Omega_\mathrm{p}^2\right)^2
-\bar
v_1a^4\Omega_\mathrm{p}^3\kappa_3\right]\nonumber\\
&+&\frac{11wa^3}{4}
\left[\left(\frac{\mathrm{d}\kappa_1}{\mathrm{d}s}\right)^2 +
\left(\frac{\mathrm{d}\kappa_2}{\mathrm{d}s}\right)^2\right],
\label{U}
\end{eqnarray}
where $\tilde u=33u/8$ and
$\kappa_0^2a^2=4(\epsilon_\mathrm{p}^2-3\epsilon_0^2)/3$. Note
that since $k_\mathrm{s}/u\rightarrow\infty$, the effect of making
$k_5$ nonzero may be captured by shifting the value of $\bar v_1$
in the model with $k_5=0$.

\begin{figure}
\includegraphics[height=2in]{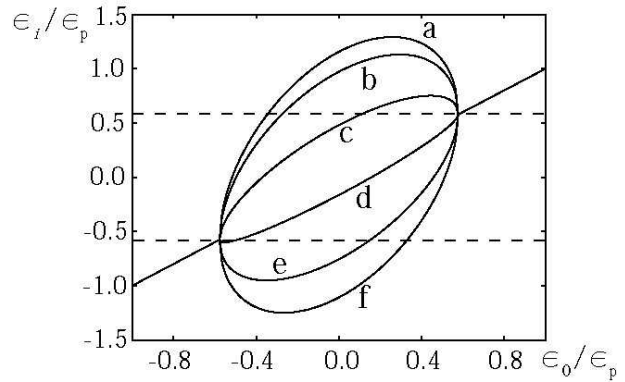}
\caption{Protofilament extension $\epsilon_i$ vs. preferred
extension $\epsilon_0$ of the inner core for $i=1$ (a), $i=2$ and
$i=11$ (b), $i=3$ and $i=10$ (c), $i=4$ and $i=9$ (d), $i=5$ and
$i=8$ (e), $i=6$ and $i=7$ (f). The region between the horizontal
dashed lines is the unstable region. } \label{protostretch}
\end{figure}

\subsection{Ground states}

To find the configuration of the filament, we minimize the energy
$\int U\mathrm{d}s $ by varying $\kappa_1$, $\kappa_2$, and
$\kappa_3$. In this subsection, we find the configuration in the
absence of external moments and forces and as a function of the
material parameters. In our discussion we will only consider
straight and helical filaments.  The cooperative term has an
absolute minimum for constant $\kappa_1$ and $\kappa_2$;
therefore, as discussed above we take $f=\pi/2$, or $\kappa_1=0$.
Minimizing $U$ over $\kappa_2$ leads to
\begin{equation}
\kappa_2=0\label{kappa2I}
\end{equation}
for $|\epsilon_0|>\epsilon_\mathrm{p}/\sqrt{3}$ (or
$\kappa_0^2<0$), and
\begin{equation}
\kappa_2^2a^2=4(\epsilon_\mathrm{p}^2-3\epsilon_0^2)/3\label{kappa2II}
\end{equation}
for $|\epsilon_0|<\epsilon_\mathrm{p}/\sqrt{3}$ (or
$\kappa_0^2>0$).

To interpret these results physically, consider a straight
filament. If $|\epsilon_0|<\epsilon_\mathrm{p}/\sqrt{3}$, then the
strain $\epsilon_i$ of each of the eleven protofilaments lies in
the unstable region of the double-well potential, where
$\partial^2U_\mathrm{s}/\partial\epsilon^2<0$. Therefore, the straight state is
unstable. The filament can lower its energy by bending, since in
the bent state, the protofilaments on the inside of the curve will
have an extension close to $-\epsilon_\mathrm{p}$, and the
protofilaments on the outside of the curve will have an extension
close to $\epsilon_\mathrm{p}$. Figure~\ref{protostretch} shows the
stretch
$\epsilon_i=\epsilon_0+(4/3)^{1/2}\sqrt{\epsilon_\mathrm{p}^2-3\epsilon_0^2}\cos[2\pi(i-1)/11]$
of the individual protofilaments as a function of $\epsilon_0$.
Note that when the filament is bent, there are always some
protofilaments with an unstable extension
$|\epsilon_i|<\sqrt{3}\epsilon_\mathrm{p}$, even though the
filament as a whole is stable.

This last observation illustrates the importance of the central
stretching potential, Eq.~(\ref{U0eqn}). If there were no central
core, then $k_\mathrm{s}=0$, as in the models
of~\cite{calladine1975} and~\cite{hasegawa_etal1998}. When
$k_\mathrm{s}=0$ in our model, both $\epsilon$ and $\kappa_2$ are
determined by minimizing $U_0+U_\mathrm{s}$. States with
$\kappa_2\ne0$ are still possible, despite the absence of lattice
mismatch, but straight states always have lower energy. This
metastability of curved states when $k_\mathrm{s}=0$ is consistent
with the experimental observation that filaments formed from
subunits with truncated ends are straight, unless the subunits
grow on a helical seed filament
~\cite{vonderviszt_aizawa_namba1991}. The cooperative interactions
at the junction between the seed filament and the filament
consisting of truncated subunits can favor curvature.

\begin{figure}
\includegraphics[height=2in]{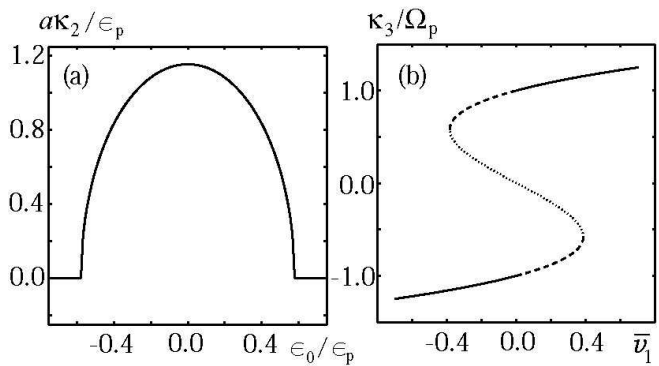}
\caption{(a) Curvature vs. $\epsilon_0$ and (b) twist vs. $\bar
v_1$ for the simple model of Section~\ref{simpmod}. In (b), the
solid lines correspond to the states of lowest energy, the dashed
lines to metastable states, and the dotted line to unstable
states.} \label{c_and_t}
\end{figure}

The twist is determined by minimizing $U$ over $\kappa_3$, which
yields
\begin{equation}
\kappa_3^3-\Omega_\mathrm{p}^2\kappa_3=\bar
v_1\Omega_\mathrm{p}^3.\label{twisteqn}
\end{equation}
When $\bar v_1^2>4/27$, there is only one value of $\kappa_3$ that
solves Eq.~(\ref{twisteqn}). This solution is stable since
$\partial^2U/\partial\kappa_3^2>0$. When $\bar v_1^2<4/27$, there
are three solutions. One of these solutions has
$\partial^2U/\partial\kappa_3^2<0$, and is unstable. Of the two
stable solutions, the one with the same sign as $\bar v_1$ has
lower energy.

Figure~\ref{c_and_t}(a) shows the dependence of curvature $\kappa_2$
on the preferred stretch $\epsilon_0$ of the central potential,
and Fig.~\ref{c_and_t}(b) shows the dependence of twist $\kappa_3$
on the asymmetry parameter $\bar v_1$. These figures directly show
the continuous nature of the transition from straight to bent, and
the discontinuous nature of the transition from left- to
right-handed. Note that since $\epsilon_0$ and $\bar v_1$ are
independent parameters, the stretch and the twist are not
coupled in our simple model.

\begin{figure}
\includegraphics[height=1.9in]{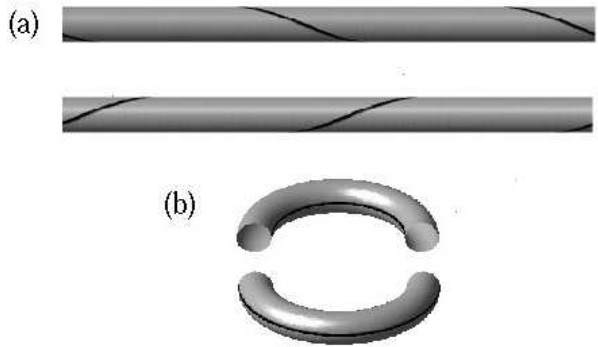}
\caption{(a) Applying twisting moments to the right-handed rod
(top) causes it to snap through to a left-handed state (bottom).
(b) The rod on top has $\kappa_2>0$; the rod on the bottom has
$\kappa_2<0$. Applying moments as shown to the top rod causes it
to snap through.} \label{snapthrough}
\end{figure}

\subsection{Response to external moment}\label{momsec}

The goal of this subsection is to calculate $\bm{\kappa}$ for a
filament subject to an external moment
$\mathbf{M}=M\hat\mathbf{z}$, where the $z$-axis is along the
helical axis for helices and along the tangent vector for straight
filaments. We begin with a qualitative discussion. Consider a
straight filament with a double-well potential for twist and a
single well potential for curvature [for example, a filament with
$\kappa_0^2<0$; see Eq.~(\ref{U})]. To simplify the discussion,
take the twist potential to be symmetric: $\bar v_1=0$. Suppose
the filament is in a state of right-handed twist when $M=0$.
Applying a positive twisting moment will tend to increase the
twist of the filament. As the positive moment increases, the twist
of the filament will increase continuously. Applying a negative
moment to the filament will tend to untwist it
[Fig.~\ref{snapthrough}(a), top]. As the magnitude of the negative
moment increases, the filament will untwist continuously until it
snaps through to a left-handed twist at a critical moment. Further
increase in the magnitude of the moment will cause the left-handed
twist to increase continuously. Once the moment is released, the
filament will remain in a left-handed state [see
Fig.~\ref{snapthrough}(a), bottom]. This state is the same energy as the
right-handed state when $\bar v_1=0$ (and $k_5=0$); when $\bar
v_1>0$, this state is metastable.

\begin{figure}
\includegraphics[height=1.4in]{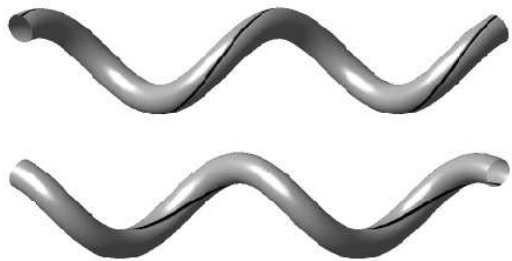}
\caption{Two helical filaments. The black line traces along the
path of the protofilament $i=1$. Top: a filament with $\kappa_2<0$
and $\kappa_3<0$.  Applying axial moments of sufficient magnitude
in the directions shown causes the filament to snap through to a
state with $\kappa_2>0$ and $\kappa_3>0$ (bottom).}
\label{twohelices}
\end{figure}

The situation is similar for a filament with a double-well
potential for curvature and a single well potential for twist (for
example, a filament with $\kappa_0^2>0$ and
$\Omega_\mathrm{p}^2<0$). In the absence of external moment, the
shape of this filament will be an arc of a circle. Suppose
$\kappa_2>0$; that is, the material frame vector
$\hat\mathbf{e}_1$ points to the center of curvature, or
equivalently, the protofilament with $i=1$ lies on the inside of
the curved surface [Fig.~\ref{snapthrough}(b), top]. Applying
bending moments with sufficient magnitudes causes the filament to snap through to
a state of negative $\kappa_2$. When the bending moments are
released, the filament remains in a state of $\kappa_2<0$.

A helical filament with double-well potentials for curvature and
twist has snap-through behavior in both the curvature and twist,
since a moment $M$ along the helical axis may be resolved into a
twisting moment $M_\mathrm{t}$ along the filament tangent vector
and a bending moment $M_\mathrm{b}$ perpendicular to the filament
tangent vector. Figure~\ref{twohelices} shows helical states
analogous to those of Fig.~\ref{snapthrough}(a) and (b).

We now turn to the quantitative determination of the response to
an external moment. The variables $\kappa_\mu$ used in the
calculation of the phase diagram are intrinsic; that is, they
depend on the shape but not the orientation of the filament. Since
an external moment specifies a direction in space, it is
convenient to use extrinsic variables such as the Euler angles,
which are defined using a space-fixed frame. We will use the
convention of Love~\cite{love1944}. The material frame at $s$ is
generated by first rotating the frame
$\{\hat\mathbf{x},\hat\mathbf{y},\hat\mathbf{z}\}$ about
$\hat\mathbf{z}$ through an angle $\psi$ to obtain the frame
$\hat\mathbf{e}'_i$, then by rotating this frame about
$\hat\mathbf{e}_2'$ by $\theta$ to obtain the frame
$\hat\mathbf{e}''_i$, and finally by rotating the frame
$\hat\mathbf{e}''_i$ by $\phi$ about $\hat\mathbf{e}''_3$ to
obtain $\hat\mathbf{e}_i$ [see Eqs.~(\ref{e1p})--(\ref{e3euler}) in
Appendix~\ref{euleranglessection}]. Our convention is to have the
Euler angles lie in the ranges
\begin{eqnarray}
0&\le&\psi<2\pi\label{psirange}\\
-\pi/2&\le&\theta\le\pi/2\label{thetarange}\\
0&\le&\phi<2\pi,\label{phirange}
\end{eqnarray}
so that the component of $\hat\mathbf{e}_3$ along $\hat\mathbf{z}$
is always nonnegative.
 The formulas for the material frame
vectors in terms of Euler angles are given in
Eqs.~(\ref{e1euler}--\ref{e3euler}). To determine the components
of the vector $\bm{\kappa}$ in terms of Euler angles, rewrite
Eq.~(\ref{onframeeqn}) as
$\kappa_\mu=\epsilon_{\mu\nu\lambda}\mathrm{d}\hat\mathbf{e}_\nu/\mathrm{d}s\cdot\hat
\mathbf{e}_\lambda$, where $\epsilon_{ijk}$ is the Levi-Civita symbol, and find
\begin{eqnarray}
\kappa_1&=&\dot\theta\sin\phi-
\dot\psi\sin\theta \cos\phi\label{k1}\\
\kappa_2 &=&\dot\theta \cos\phi
+\dot\psi\sin\theta\sin\phi\label{k2}\\
\kappa_3&=&\dot\phi + \dot\psi\cos\theta,\label{k3}
\end{eqnarray}
where the dot signifies differentiation with respect to $s$; for
example, $\dot\theta=\mathrm{d}\theta/\mathrm{d} s$.

In equilibrium, equal and opposite moments must be applied to the
ends of the rod. The principle of virtual work is therefore
\begin{equation}
\delta
E-\mathbf{M}\cdot[\delta\bm{\Upsilon}(L)-\delta\bm{\Upsilon}(0)]=0,
\end{equation}
where $\delta\Upsilon_\mu(s)$ is the angle of the virtual rotation
of the material frame about $\hat\mathbf{e}_\mu$. For example, the
angle of a virtual rotation of the material frame about
$\hat\mathbf{e}_3$ is given by the $\hat\mathbf{e}_2$ component of
the infinitesimal change in $\hat\mathbf{e}_1$:
$\delta\Upsilon_3=\hat\mathbf{e}_2\cdot \delta\hat\mathbf{e}_1$.
The general relation follows by taking cyclic permutations,
$\delta\Upsilon_\mu
=\epsilon_{\mu\nu\lambda}\delta\hat\mathrm{e}_\nu
\cdot\hat\mathrm{e}_\lambda$. Using
Eqs.~(\ref{e1euler}--\ref{e3euler}) to calculate
\begin{eqnarray}
\delta\Upsilon_1&=&-\delta\psi\sin\theta\cos\phi+\delta\theta\sin\phi
\label{dU1}\\
\delta\Upsilon_2&=&\delta\psi\sin\theta\sin\phi+\delta\theta\cos\phi
\label{dU2}\\
\delta\Upsilon_3&=&\delta\psi\cos\theta+\delta\phi \label{dU3}
\end{eqnarray}
yields
$\mathbf{M}\cdot\delta\bm{\Upsilon}=M\delta\psi+M\cos\theta\delta
\phi$.

Taking the variation of $E=\int U\mathrm{d}s $ with respect to
$(\Xi_1,\Xi_2,\Xi_3)=(\psi,\theta,\phi)$ leads to the
Euler-Lagrange equations
\begin{equation}
\frac{\partial
U}{\partial\Xi_a}-\frac{\mathrm{d}}{\mathrm{d}s}\frac{\partial
U}{\partial\dot\Xi_a}+\frac{\mathrm{d}^2}{\mathrm{d}s^2}
\frac{\partial U}{\partial\ddot\Xi_a}=0,\label{eulerlagrange}
\end{equation}
where $a$ runs from 1 to 3. Since $U$ is independent of $\psi$,
$\phi$, and $\ddot\phi$ (see the appendix), these equations
simplify to
\begin{eqnarray}
-\frac{\mathrm{d}}{\mathrm{d}s}\frac{\partial
U}{\partial\dot\psi}+\frac{\mathrm{d}^2}{\mathrm{d}s^2}
\frac{\partial U}{\partial\ddot\psi}&=&0\label{el1}\\
\frac{\partial
U}{\partial\theta}-\frac{\mathrm{d}}{\mathrm{d}s}\frac{\partial
U}{\partial\dot\theta}+\frac{\mathrm{d}^2}{\mathrm{d}s^2}
\frac{\partial U}{\partial\ddot\theta}&=&0\label{el2}\\
-\frac{\mathrm{d}}{\mathrm{d}s}\frac{\partial
U}{\partial\dot\phi}&=&0\label{el3}
\end{eqnarray}

To complete the description of the variational problem, we assume
natural boundary conditions: $\delta\Xi_a(0)$ and $\delta\Xi_a(L)$
are arbitrary, and $\delta\dot\Xi_a(0)= \delta\dot\Xi_a(L)=0$.
Noting again that $U$ is independent of $\ddot\phi$, we find that
these natural boundary conditions imply
\begin{eqnarray}
\left[\frac{\partial U}{\partial
\ddot\Xi_{a}}\right]_{s=L}&=&0\label{bcss}\\
\left[\frac{\partial U}{\partial\dot\psi} -
\frac{\mathrm{d}}{\mathrm{d}s}\frac{\partial
U}{\partial\ddot\psi}\right]_{s=L}&=&M\label{bcpsis}\\
\left[\frac{\partial U}{\partial\dot\theta} -\frac{\mathrm
d}{\mathrm {d}s} \frac{\partial
U}{\partial\ddot\theta}\right]_{s=L} &=& 0\label{bcthetas}\\
\left[\frac{\partial U}{\partial\dot\phi}  \right]_{s=L}
&=&M\cos\theta.\label{bcphis}
\end{eqnarray}
Exactly the same conditions apply at $s=0$.

The equations~(\ref{el1}--\ref{bcphis}) are derived without making
any assumptions on the form of $\psi$, $\theta$, and $\phi$. We
now turn to the special cases of straight and helical filaments. A
helix has a parameterization
\begin{equation}
\mathbf{r}(s)=\left(\frac{\sin\theta}{\dot\psi}\sin(\dot\psi s),
-\frac{\sin\theta}{\dot\psi}\cos(\dot\psi s),s\cos\theta\right),
\label{r-eulerangle}
\end{equation}
with $\dot\theta=0$ and $\ddot\psi=0$. For constant $\theta$ and
constant $\dot\psi$, Eq.~(\ref{bcss}) with $a=2$ (and the
corresponding equation at $s=0$) imply $\partial
U/\partial\ddot\theta=(11/2)wa^3\dot\psi\dot\phi\sin\theta=0$ at
$s=L$ and $s=0$. But $\theta\neq0$ and $\dot\psi\neq0$ for a
helix; therefore $\dot\phi=0$ at the ends of the filament. On the
other hand, since $U$ is independent of $\phi$ and $\ddot\phi$,
Eq.~(\ref{el3}) implies
\begin{equation} \frac{\partial
U}{\partial \dot\phi}=va^4\left(\dot\phi+\dot\psi\cos\theta\right)
\left[\left(
\dot\phi+\dot\psi\cos\theta\right)^2-\Omega_\mathrm{p}^2\right]\label{dudphi-dot}
\end{equation}
is constant. Since $\dot\psi$ and $\theta$ are both constant,
$\dot\phi$ must be constant. Therefore, $\dot\phi=0$ for all $s$.
Note that without the cooperative term, this conclusion would not
follow. With loss of generality we choose $\phi=\pi/2$. A shift in
$\phi$ amounts to redefining which protofilament has $i=1$.

In general, the Euler angle $\phi$ differs from the angle $f$
defined in Fig.~\ref{fsmatfig}. However, there is a simple
relation between $\phi$ and $f$ for a helical rod. To derive this
relation, use Eqs.~(\ref{e1pp}--\ref{e3pp}) to show
\begin{equation}
\kappa\hat\mathbf{n}=\mathrm{d}\hat\mathbf{e}''_3/\mathrm{d}
s=\dot\theta\hat\mathbf{e}''_1+
\dot\psi\sin\theta\hat\mathbf{e}''_2\label{e2n}
\end{equation}
Thus, for a helix for which $\dot\theta=0$, the normal is parallel
or antiparallel to $\hat\mathbf{e}_2''$;
$\hat\mathbf{n}=\hat\mathbf{e}_2''$ and $f=\phi$ when
$\kappa_2>0$, and $\hat\mathbf{n}=-\hat\mathbf{e}_2''$ and
$f=\phi+\pi$ when $\kappa_2<0$. With our choice for $\phi$,
$f=\pi/2$ when $\kappa_2>0$ and $f=3\pi/2$ when $\kappa_2<0$.

Returning to the Euler-Lagrange equations, we note that
Eqs.~(\ref{el3}) and~(\ref{bcphis}) imply $\partial
U/\partial\dot\phi=M\cos\theta$ for all $s$. Likewise, since
$\partial U/\partial\ddot\psi=0$ when $\dot\theta=\ddot\psi=0$,
Eqs.~(\ref{el1}) and~(\ref{bcpsis}) imply $\partial
U/\partial\dot\psi=M$ for all $s$. The other Euler-Lagrange
equation, Eq.~(\ref{el2}) with $\dot\theta=\ddot\psi=\dot\phi=0$,
does not yield an independent equation. Thus,
\begin{eqnarray}
\dot\psi\sin\theta\left(\dot\psi^2\sin^2\theta-\kappa_0^2\right)&=&
\frac{M}{\tilde ua^4}\sin\theta\label{eaeq1}\\
\dot\psi\cos\theta\left(\dot\psi^2\cos^2\theta-\Omega^2_\mathrm{p}\right)&=&
\frac{M}{va^4}\cos\theta.\label{eaeq2}
\end{eqnarray}

Note that under the assumption of uniform $\kappa_1$ and
$\kappa_2$, Eqs.~(\ref{eaeq1}--\ref{eaeq2}) also follow from the
constitutive relations
\begin{equation}
M_\mu=\frac{\partial U}{\partial\kappa_\mu}\label{Mkappa},
\end{equation}
with $M_1=0$, $M_2=M\sin\theta$, and $M_3=M\cos\theta$.

\begin{figure}
\includegraphics[height=1.4in]{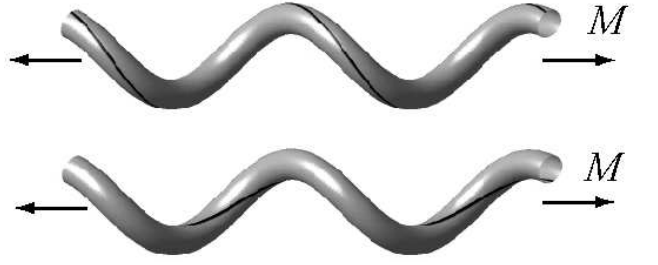}
\caption{Two helices with the same curvature and twist, one with
$\kappa_2<0$ (top), and one with $\kappa_2>0$ (bottom). In each
case, the black line traces the path of the protofilament $i=1$.}
\label{twohelicesII}
\end{figure}

First we analyze a rod that is helical when $M=0$; hence
$\kappa_0^2>0$, and we may assume $\theta\neq0$ and
$\theta\ne\pi/2$. It is convenient to define
$\sigma^2=\kappa_0^2+\Omega_\mathrm{p}^2$, the pitch angle
$\theta_0$ for zero applied moment,
$\tan\theta_0=\kappa_0/\Omega_\mathrm{p}$, and the dimensionless
moment $m=M(\tilde u+v)/(\tilde u v a^4\sigma^3)$. The angle
$\theta$ may be eliminated from Eqs.~(\ref{eaeq1})
and~(\ref{eaeq2}) to yield a simple cubic equation for $\dot\psi$:
\begin{equation}
\dot\psi^3/\sigma^3-\dot\psi/\sigma=m.\label{hpsieqn}
\end{equation}
The curve relating $\dot\psi/\sigma$ to $m$ is precisely the same
as the curve in Fig~\ref{c_and_t}(b). Equation~\ref{hpsieqn}
determines the handedness of the filament since
$\kappa_3=\dot\psi\cos\theta$, and $\cos\theta>0$. Combining
Eq.~(\ref{hpsieqn}) with Eq.~(\ref{eaeq1}) leads to
\begin{equation}
\sin^2\theta=\frac{v}{\tilde
u+v}+\left(\sin^2\theta_0-\frac{v}{\tilde
u+v}\right)\frac{\sigma^2}{\dot\psi^2}.\label{hthetaeqn}
\end{equation}
The sign of $\kappa_2=\dot\psi\sin\theta$ is not determined by
these equations since Eq.~(\ref{hthetaeqn}) is invariant under
$\theta\mapsto-\theta$. The consequence of this invariance is that
for every $M$, there are two helical solutions with the same
curvature and twist, one with $\kappa_2<0$ (as in
Fig.~\ref{twohelicesII}, top), and one with the $\kappa_2>0$ (as
in Fig.~\ref{twohelicesII}, bottom). This symmetry remains valid
in the full model also.

Under the restrictions of uniform $\dot\psi$, uniform $\theta$,
and $\phi=\pi/2$, these solutions are stable if and only if the
Hessian matrix
\begin{equation}
\mathsf{H}=\left[\matrix{\partial^2U/\partial\dot\psi^2
&\partial^2U/(\partial\dot\psi\partial\theta)\cr
\partial^2U/(\partial\theta\partial\dot\psi)&\partial^2U/
\partial\theta^2}\right]\label{Hessian_ab}
\end{equation}
is positive definite, or equivalently,
$\mathrm{d}\dot\psi/\mathrm{d}M>0$. To see why these conditions
are equivalent, note under the restrictions just mentioned the
Euler-Lagrange equations are
\begin{eqnarray}
\partial U/\partial\dot\psi-M&=&0\label{ELpsi}\\
\partial U/\partial\theta&=&0.\label{ELtheta}
\end{eqnarray}
Differentiating these equations with respect to $M$ and solving
leads to
\begin{equation}
\frac{\mathrm{d}\dot\psi}{\mathrm{d}M}=H_{22}/\det
\mathsf{H}\label{dpsidm}
\end{equation}

\begin{figure}
\includegraphics[height=1.95in]{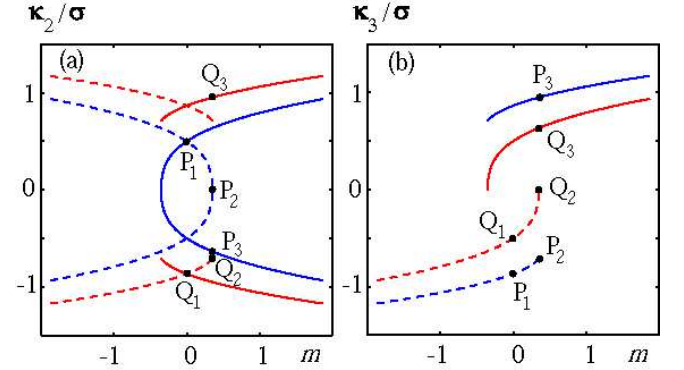}
\caption{(a) Signed curvature $\kappa_2$ vs. $m$ for $\tilde u=v$
and $\theta_0=\pi/6$ (blue curves) and $\theta_0=\pi/3$ (red
curves). The dashed lines correspond to left-handed states. (b)
Twist $\kappa_3$ vs. $m$ with parameters as in (a).}
\label{k2k3Ifig}
\end{figure}

The matrix element $H_{22}=2(\tilde
u+v)a^4\dot\psi^4\sin^2\theta\cos^2\theta>0$ as long as
$\theta\ne0$ and $\theta\ne\pi/2$. Therefore, the solution is
stable when $\mathrm{d}\dot\psi/\mathrm{d}M>0$, or
$\dot\psi^2>\sigma^2/3$, and unstable when
$\mathrm{d}\dot\psi/\mathrm{d}M<0$, or $\dot\psi^2<\sigma^2/3$. An
alternate route to this conclusion is to use Eq.~(\ref{hthetaeqn})
to eliminate $\theta$ from the total potential energy per length
$\bar U=U-M\dot\psi$. The energy $\bar U$ in terms of $\dot\psi$
is a simple double-well potential with bias given by $m$:
\begin{equation}
\bar U=a^4\sigma^4\frac{\tilde uv}{\tilde
u+v}\left[\frac{1}{4}\left(\frac{\dot\psi^2}{\sigma^2} -
1\right)^2 - m \frac{\dot\psi}{\sigma}\right].\label{esimpleI}
\end{equation}

We now examine the behavior of the solutions. Suppose the helix is
left-handed with $\kappa_2>0$ for $m=0$ [P$_1$ in
Fig.~\ref{k2k3Ifig}(a) and (b)]. Eq.~(\ref{hthetaeqn}) has no solution
for $\theta$ when $\dot\psi^2<\dot\psi^2_\mathrm{c}$, where
\begin{equation}
\dot\psi^2_\mathrm{c}=\pm\sigma^2\left(1-\frac{\tilde
u+v}{v}\sin^2\theta_0\right)\label{psic}
\end{equation}
for $\sin^2\theta_0\lessgtr v/(\tilde u+v)$. When $\sin^2\theta_0<
v/(\tilde u+v)$, $\theta$ decreases to a minimum value of $0$ as
$\dot\psi$ decreases to $\dot\psi_\mathrm{c}$; otherwise, $\theta$
increases to a maximum value of $\pi/2$ as $\dot\psi$ decreases to
$\dot\psi_\mathrm{c}$. If $\dot\psi_\mathrm{c}^2>\sigma^2/3$, all
these solutions are stable. This condition places a further
constraint on $\theta_0$, leading to two regimes. In the regime
with $0<\sin^2\theta_0<(2/3)v/(\tilde u+v)$, an applied moment
$m>0$ causes the helix to deform continuously into a
\textit{straight} twisted filament at $m=m_\mathrm{c}$ (P$_2$ in
Fig.~\ref{k2k3Ifig}); as the moment increases beyond
$m_\mathrm{c}$, there is a discontinuous transition to a
right-handed helical filament (P$_3$ in Fig~\ref{k2k3Ifig}). In
the regime with $(v+\tilde u/3)/(\tilde u+v)<\sin^2\theta_0<1$, an
applied moment $m>0$ causes the helix to deform continuously into
an open coil with no twist at $m=m_\mathrm{c}$ (Q$_2$ in
Fig.~\ref{k2k3Ifig}); as the moment increases beyond
$m_\mathrm{c}$, there is a discontinuous transition to a
right-handed helical filament (Q$_3$ in Fig.~\ref{k2k3Ifig}). The
critical dimensionless moment is
\begin{equation}
m_\mathrm{c}=\frac{\tilde u+v}{v}\sin^2\theta_0
\sqrt{\pm\left(1-\frac{\tilde u+v}{v}\sin^2\theta_0\right)}
\label{mc}
\end{equation}
(again for $\sin^2\theta_0\lessgtr v/(\tilde u+v)$).

\begin{figure}
\includegraphics[height=1.95in]{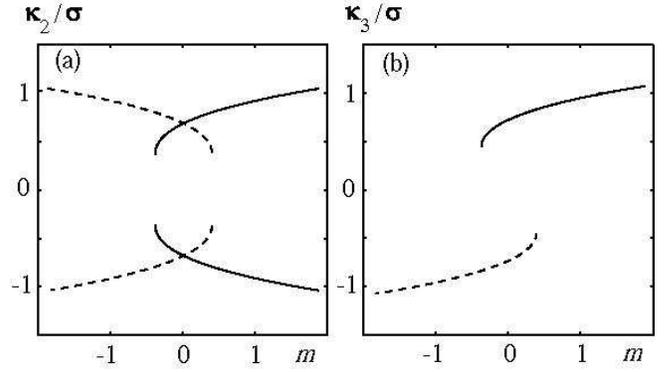}
\caption{(a) Signed curvature $\kappa_2$ vs. $m$ for $\tilde u=v$
and $\theta_0=43^\circ\approx0.2389\pi$ . The dashed lines
correspond to left-handed states. (b) Twist $\kappa_3$ vs. $m$
with parameters as in (a).} \label{k2k3IIfig}
\end{figure}

If $\dot\psi_\mathrm{c}^2<\sigma^2/3$, or equivalently,
$(2/3)v/(\tilde u+v)<\sin^2\theta_0<(v+\tilde u/3)/(\tilde u+v)$, then as
$m$ increases, the helix becomes unstable before either the
straight or coiled state is reached. Thus, there is a
discontinuous transition directly from a left-handed helix to a
right-handed helix at $m=m_\mathrm{c}=2/(3\sqrt{3})$
(Fig~\ref{k2k3IIfig}).

The solution curves of Figs.~\ref{k2k3Ifig} and \ref{k2k3IIfig}
are symmetric under the combined operation $m\mapsto-m$ and
$\kappa_3\mapsto-\kappa_3$, since the nonlinear twist potential is
even in $\kappa_3$ in the simple model with $\bar v_1=0$. The full
model has $\bar v_1\ne0$ and therefore will not have this
symmetry.

Now suppose the filament is straight when $M=0$. In this case
$\kappa_0^2<0$, and there is a branch of solutions corresponding
to straight filaments with $\theta=0$ and $\dot\psi$ given by
\begin{equation}
\dot\psi^3/\sigma^3-(\dot\psi/\sigma)\cos^2\theta_0=m\tilde
u/(\tilde u+ v)\label{psieqnst}
\end{equation}

\begin{figure}
\includegraphics[height=1.95in]{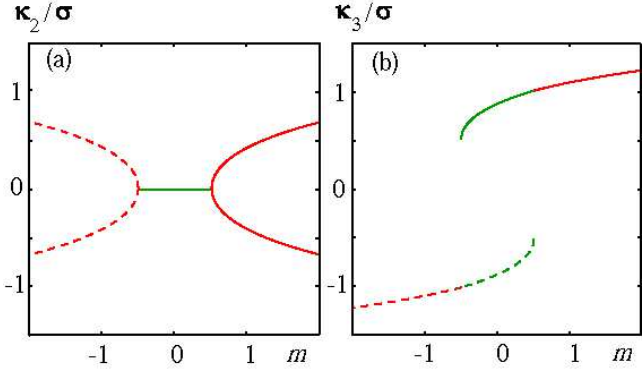}
\caption{(a) Signed curvature $\kappa_2$ vs. $m$ for $\tilde u=v$
and $\theta_0=\pi/6$. The dashed lines correspond to left-handed
states. The red curves correspond to helical solutions; the green
curves correspond to straight solutions. (b) Twist $\kappa_3$ vs.
$m$ with parameters as in (a).} \label{k2k3stfig}
\end{figure}

The Hessian matrix for the straight solutions is diagonal, with
\begin{eqnarray}
H_{11}&=&va^4(3\dot\psi^2-\sigma^2\cos^2\theta_0) \label{H11st}\\
H_{22}&=&a^4\dot\psi^2(\tilde u\sigma^2\sin^2\theta_0 +
v\sigma^2\cos^2\theta_0 - v\dot\psi^2). \label{H22st}
\end{eqnarray}
Thus, the straight solutions are stable for
$\dot\psi<\dot\psi_\mathrm{h}$, where
\begin{equation}
\dot\psi_\mathrm{h}^2=\sigma^2\left(1+\frac{\tilde
u-v}{v}\sin^2\theta_0\right). \label{psicII}
\end{equation}
Equivalently, the straight states are stable for $m<m_\mathrm{h}$,
where
\begin{equation}
m_\mathrm{h}=\frac{\tilde
u+v}{v}\sin^2\theta_0\sqrt{1+\frac{\tilde
u-v}{v}\sin^2\theta_0}.\label{mcII}
\end{equation}
The green curves of Fig.~\ref{k2k3stfig} show $\kappa_2$ and
$\kappa_3$ as a function of $m$.

Once $|m|$ is sufficiently large, there is also a branch of
helical solutions, with $\dot\psi$ given by
\begin{equation}
\dot\psi^3/\sigma^3-\cos(2\theta_0)\dot\psi/\sigma=m,\label{shpsieqn}
\end{equation}
and $\theta$ given by
\begin{equation}
\sin^2\theta=\frac{v}{\tilde u+v}-\frac{v\cos^2\theta_0+\tilde
u\sin^2\theta_0 }{\tilde
u+v}\frac{\sigma^2}{\dot\psi^2}.\label{sthetaeqn}
\end{equation}
The same stability analysis described earlier,
Eqs.~(\ref{ELpsi}--\ref{dpsidm}), implies that these helical
solutions are stable if and only if
$\mathrm{d}\dot\psi/\mathrm{d}M>0$, or
$\dot\psi^2>(\sigma^2/3)\cos(2\theta_0)$. Using
Eq.~(\ref{sthetaeqn}) to eliminate $\theta$ from the energy leads
to the potential per unit length
\begin{equation}
\bar U=a^4\sigma^4\left[\frac{1}{4}\tilde u \sin^4\theta_0+
\frac{1}{4}v\left(\frac{\dot\psi^2}{\sigma^2}-1\right)^2-\frac{\tilde
u v}{\tilde u+v}m\frac{\dot\psi}{\sigma}\right].\label{Est}
\end{equation}
To determine the critical $m$ at which helical solutions are
possible, note that Eq.~(\ref{sthetaeqn}) has no solution for
$\theta$ when $\psi^2<\psi^2_\mathrm{h}$. Therefore, the branch of
helical solutions begins when $\psi=\psi_\mathrm{h}$ and
$m=m_\mathrm{h}$, precisely where the straight states become
unstable. The red curves of Fig.~\ref{k2k3stfig} show $\kappa_2$
and $\kappa_3$ for $\tilde u=v$ and $\theta_0=\pi/6$. Note that
since $\dot\psi_\mathrm{h}^2>(\sigma^2/3)\cos(2\theta_0)$, the
branch of helical solutions is stable.

Care must be taken in the interpretation the graphs of
Fig.~\ref{k2k3stfig}.  A straight filament under external moment
can destabilize via the familiar twist instability of an ordinary
elastic rod~\cite{landau_lifshitz_elas}, which is different from
the polymorphic transitions studied in this article. For a rod
with ends that are not clamped, the critical moment for this
instability is $M_\mathrm{twist}=2\pi A/L$, where $A$ is the
bending stiffness and $L$ is the length. For our rod, $A$ is the
coefficient of $\kappa_2^2$ in the energy; that is,
$A=a^4\sigma^2u\sin^2\theta_0$. Thus, we may estimate the length
$L_\mathrm{h}$ at which a straight rod subject the moment
$m_\mathrm{h}$ just sufficient to allow helical solutions will
undergo the twisting instability:
\begin{equation}
L_\mathrm{h}=2\pi\sqrt{v/(v\Omega_\mathrm{p}^2+u\kappa_0^2).}\label{L0}
\end{equation}
Since $\kappa_0$ and $\Omega_\mathrm{p}$ are of the order of
microns, $L_\mathrm{h}$ is about 6 microns.  When
$L>L_\mathrm{h}$, we expect a straight flagellar filament will
buckle when subject to a dimensionless moment less than
$m_\mathrm{h}$.

\subsection{Response to external force}\label{forcesec}
\begin{figure}
\includegraphics[height=0.9in]{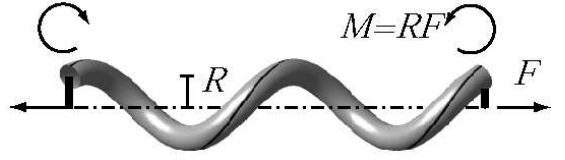}
\caption{The line of action of the force coincides with the axis
of the helix, leading to moments applied at each end.}
\label{twohelices1}
\end{figure}

With the simplified limit of a stiff central spring still valid,
we study the response of the filament to an applied external force
$F\hat{\mathbf z}$. As mentioned earlier, we only consider
filament shapes which are either straight or helical.  The line of
action of the force coincides with the axis of the helix, as if
there were rigid bars perpendicular to $\hat\mathbf{z}$ at each
end of the helix (Fig.~\ref{twohelices1}). Thus, the applied force
also leads to an applied moment. If $\mathbf{r}(0)$ is the origin
for the moment, then the applied moments at the ends of the rod
are $\mathbf{r}(0)\times F\hat\mathbf{z}=-RF\hat{\bm\varphi}(0)$
at $s=0$ and $RF\hat{\bm\varphi}(0)$ at $s=L$, where $\varphi$
(not to be confused with the Euler angle $\phi$) is the azimuthal
coordinate and $R$ is the radius of the helix. To see that this
loading condition is compatible with helical shapes, consider the
equilibrium equations
\begin{eqnarray}
\dot{\mathbf M}+{\mathbf r}\times{\mathbf F}&=&{\mathbf 0}
\label{Mbal}\\
\dot{\mathbf F}&=&{\mathbf 0},\label{Fbal}
\end{eqnarray}
where ${\mathbf M}(s)$ is the moment due to internal stresses
acting on the rod through the cross-section at $s$, and ${\mathbf
F}(s)$ is the force acting on the rod through the cross-section at
$s$~\cite{landau_lifshitz_elas}. Since ${\mathbf F}=F\hat{\mathbf
z}$ is constant, we can integrate the moment balance
equation~\ref{Mbal} to find ${\mathbf M}=-{\mathbf
r}(s)\times{\mathbf F}+ {\mathbf C}$, where ${\mathbf C}$ is an
integration constant. Imposing the boundary condition ${\mathbf
M}(0)=-\mathbf{r}(0)\times\mathbf{F}=RF\hat{\bm\varphi}(0)$ yields
${\mathbf M}=RF\hat{\bm \varphi}(s)$ for a helix of radius $R$.
Since the moment has a constant magnitude, the constitutive
relations~(\ref{Mkappa}) imply that helical solutions with constant
curvature and torsion are possible. If the line of action of the
external force passes through the end of the rod rather than the
$z$ axis, then $|{\mathbf M}|$ is not uniform, and the curvature
and twist are not constant.

The principle of virtual work for our loading condition is
\begin{eqnarray}
\delta E&-&\mathbf{F}\cdot\left[\delta\mathbf{r}(L) -
\delta\mathbf{r}(0)\right]\nonumber\\ &-&
\left[\mathbf{M}(L)\cdot\delta{\bm\Upsilon}(L) -
\mathbf{M}(0)\cdot\delta{\bm\Upsilon}(0)\right] =0, \label{vwF}
\end{eqnarray}
where $\mathbf{M}(s)=RF\hat{\bm\varphi}(s)$ and
$\delta\Upsilon_\mu = \epsilon_{\mu\nu\lambda}\delta\hat{\mathbf
e}_\nu\cdot\hat {\mathbf e}_\lambda$ as defined before. To express
Eq.~(\ref{vwF}) in terms of the variables $\psi$, $\theta$, and
$\phi$, first use
$\hat\mathbf{z}\cdot[\mathbf{r}(s)-\mathbf{r}(0)]=
\int^L_0\hat\mathbf{e}_3\cdot\hat\mathbf{z}\mathrm{d} s$ to write
\begin{equation}
-\mathbf{F}\cdot\left[\delta\mathbf{r}(L) -
\delta\mathbf{r}(0)\right]=F\int^L_0\sin\theta
\delta\theta\mathrm{d}s.\label{Fdr}
\end{equation}
Then use the relation $\hat{\bm\varphi}=-\cos\theta\hat{\mathbf
e}_2 +\sin\theta\hat{\mathbf e}_3$ and Eqs.~(\ref{dU2})
and~(\ref{dU3}) to write $\mathbf{M}\cdot\delta{\bm\Upsilon}$ in
terms of $\psi$, $\theta$, and $\phi$. Just as in the case of an
applied external moment, the Euler-Lagrange equations eventually
imply $\dot\phi=0$, and we take $\phi=\pi/2$. Thus,
\begin{equation}
\left[\mathbf{M}(L)\cdot\delta{\bm\Upsilon}(L) -
\mathbf{M}(0)\cdot\delta{\bm\Upsilon}(0)\right]
=\left[F\frac{\sin^2\theta}{\dot\psi}\right]^L_0.
\end{equation}
Carrying out steps analogous to those that led to
Eqs.~(\ref{el1}--\ref{bcphis}) and specializing to the case of a
helix leads to
\begin{eqnarray}
\dot\psi^2\left(\dot\psi^2\sin^2\theta-\kappa_0^2\right)&=&
-\frac{F}{\tilde ua^4}\cos\theta\label{eaeq1_force}\\
\dot\psi^2\cos\theta\left(\dot\psi^2\cos^2\theta-\Omega^2_\mathrm{p}\right)&=&
\frac{F}{va^4}\sin^2\theta.\label{eaeq2_force}
\end{eqnarray}
Note that these equations also follow directly from the
constitutive relations of Eq.~(\ref{Mkappa}), $M_1=0$,
$M_2=-RF\cos\theta$, and $M_3=RF\sin\theta$.

\begin{figure}
\includegraphics[height=1.8in]{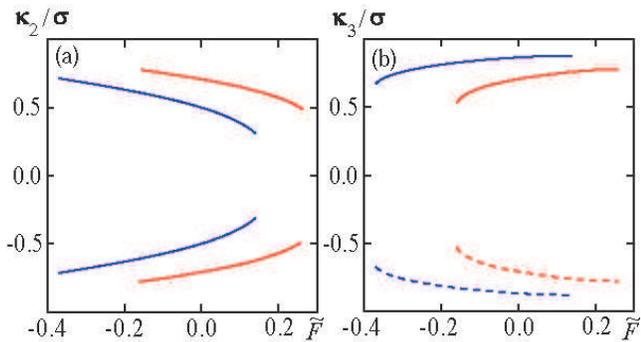}
\caption{(a) Signed curvature $\kappa_2$ vs. $\tilde F$ for $\tilde u=v$,
and $\theta_0=\pi/6$ (blue curves) and $\theta_0=\pi/4$ (red curves). 
The dashed lines correspond to left-handed states. 
(b) Twist $\kappa_3$ vs. $\tilde F$ with parameters
as in (a).} \label{forcesimplfig}
\end{figure}

We were unable to find an analytic solution to
Eqs.~(\ref{eaeq1_force}) and~(\ref{eaeq2_force}). Instead, we used
a numerical continuation method to generate the solutions
$\dot\psi$ and $\theta$ as a function of $F$~\cite{keller1977}.
The advantage of the continuation method is that it can negotiate
turning points such as those in the graph of $\kappa_3$ vs. $\bar
v_1$ in Fig.~\ref{c_and_t}. The results of our numerical
calculation are shown in Fig.~\ref{forcesimplfig}, where we have
introduced the dimensionless force $\tilde
{F}=F/(\tilde u a^4\sigma^4)$.
Figure~\ref{forcesimplfig} shows that uniform helical solutions
only exist in a finite range of $\tilde F$. To see that
Eqs.~(\ref{eaeq1_force}) and~(\ref{eaeq2_force}) do not have
solutions for large $F$, consider their asymptotic form at large
$F$:
\begin{eqnarray}
\dot\psi^4&\sim&
-\frac{F}{\tilde ua^4}\cos\theta\label{asympt1}\\
\dot\psi^4\cos\theta&\sim&
\frac{F}{va^4}\sin^2\theta.\label{asympt2}
\end{eqnarray}
The ratio of these two equations simplifies to $\tan^2\theta\sim-
v/\tilde u$, which is impossible since $\tilde u>0$ and $v>0$. We
speculate that the filament takes a non-uniform shape with varying
curvature and twist for values of $\tilde F$ outside the allowed
range of Fig.~\ref{forcesimplfig}. Note that for a given force 
$\tilde F$ and a given signed curvature $\kappa_2$, there are both 
left-handed and right-handed filaments of same magnitude of twist owing to 
symmetry of the energy function in the simplified version of the model; 
therefore the solid and dashed lines overlap in Fig.~\ref{forcesimplfig}(a).

We do not discuss the case of an otherwise \textit{straight}
filament subject to external force, since in this case we found no
examples of transitions from straight to helical states.

\section{Full model} \label{fullmod} In this section, we consider
the full model, which accounts for the inextensibility of the
filament and the lateral bonds between protofilaments. We follow
the same analysis used for the simple model. The twist-stretch
coupling arising from lateral bonds, Eq.~(\ref{U5}), makes the
phase diagram for ground states in the full model qualitatively
different from the phase diagram for ground states in the simple
model (Fig. 2 of~\cite{SrigirirajuPowers2005}). However, the
response of the filament to external moment and external force in
the full model is very similar to the response in the simple
model.

\subsection{Ground states} 
\begin{figure}
\includegraphics[height=2.5in]{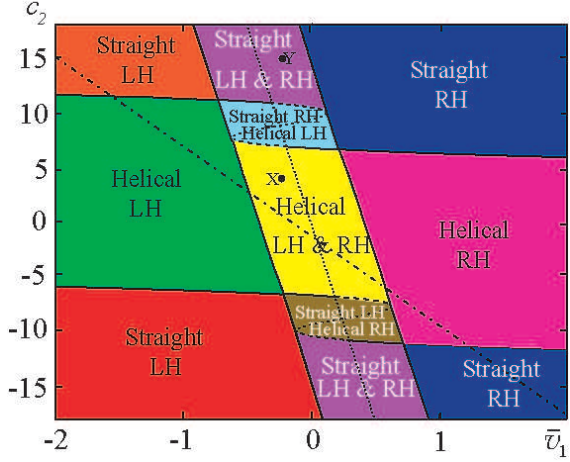}
\caption{Phase diagram for $\gamma=1.8$, $\tilde u=v$, $\bar
k_\mathrm{s}=8$, $\bar k_5=1$ and $\bar u_1=-0.5$; $\bar v_1$ and $c_2$
are dimensionless. 
The inclined dotted line defines the boundary where the energy of
the left-handed state equals the energy of the right-handed state.
The diagonal dash-dot line traces out the values of $\bar v_1$ and
$c_2$ used to calculate the curvature vs. twist plot of
Fig.~\ref{cvtfig}. The points X and Y in the metastable regions
represent the corresponding parameters used for evaluating the
response of a helical and straight filament to external loading
respectively.} \label{pdfig}
\end{figure}

The full model includes all the terms
of the energy~(\ref{E}), which we write as $E=\int V\mathrm{d}s$,
where $V=U_0+U_\mathrm{s}+U_\mathrm{t}+U_5+U_\mathrm{c}$ and we
note again that to a good approximation,
$\mathrm{d}S\approx\mathrm{d}s$. As in the analysis of the simple
model, the cooperative term is minimized when $\kappa_1=0$.
Assuming helical or straight states and minimizing $V$ over
$\epsilon$, $\kappa_2$, and $\kappa_3$, we find
\begin{eqnarray}
\frac{\epsilon^3}{\epsilon_\mathrm{p}^3}+
\frac{c_1\epsilon}{\epsilon_\mathrm{p}}-c_2-\frac{\bar k_5
\gamma\kappa_3}{\Omega_\mathrm{p}}+
\frac{3\gamma^2\epsilon\kappa_2^2}{2\epsilon_\mathrm{p}\Omega_\mathrm{p}^2}&=&0\label{Exi}\\
\frac{3\epsilon^2\kappa_2}{2\epsilon_\mathrm{p}^2\Omega_\mathrm{p}}
-\frac{\kappa_2}{2\Omega_\mathrm{p}}+
\frac{3\gamma^2\kappa_2^3}{8\Omega_\mathrm{p}^3}&=&0\label{Ekappa2}\\
\frac{v}{11u}\gamma^3
\left(\frac{\kappa_3^3}{\Omega_\mathrm{p}^3}-\frac{\kappa_3}{\Omega_\mathrm{p}}-
\bar v_1\right)-\frac{\bar
k_5\epsilon}{\epsilon_\mathrm{p}}&=&0,\label{Ekappa3}
\end{eqnarray}
where $c_1=2\bar k_\mathrm{s}-1$, $\bar
k_\mathrm{s}=k_\mathrm{s}/(22u\epsilon_\mathrm{p}^2)$, $c_2=2\bar
k_\mathrm{s}\epsilon_0/\epsilon_\mathrm{p}+\bar u_1$, 
$\bar u_1=u_1/(u\epsilon_\mathrm{p})$, $\bar
k_5=k_5/(11u\epsilon_\mathrm{p}^2)$, and
$\gamma=a\Omega_\mathrm{p}/\epsilon_\mathrm{p}$. In this section,
we choose representative values $\gamma=1.8$, $\tilde u=v$, $\bar
k_5=1$ and $\bar u_1=-0.5$ to illustrate the predictions of our model.
When $k_5\ne0$, there will be no discontinuous transitions between
helices of opposite handedness if $\bar k_\mathrm{s}$ is too
small; hence, we choose the relatively large value $\bar
k_\mathrm{s}=8$. The two parameters $\bar v_1$ and $c_2$ are the
molecular switches of the full model, just as $\bar v_1$ and
$\epsilon_0/\epsilon_\mathrm{p}$ are the switches of the simple
model: positive $\bar v_1$ favors right-handed twist, and positive
$c_2$ favors longer protofilaments. Equation~(\ref{Ekappa2}) shows
that there are straight solutions with $\kappa_2=0$ and curved
solutions with $\kappa_2\ne0$, just as in the simple model.
Evaluation of the Hessian using the independent variables
$\epsilon$, $\kappa_3$, and $\kappa_2$ for the straight states
shows that they are unstable for
$\epsilon^2<\epsilon_\mathrm{p}^2/3$. Solving Eqs.~(\ref{Exi})
and~(\ref{Ekappa3}) for $\bar v_1$ and $c_2$ when $\kappa_2=0$ and
$\epsilon=\pm\epsilon_\mathrm{p}/\sqrt{3}$ yields the boundaries
between the straight and the curved states shown in the phase
diagram of Fig.~\ref{pdfig}. The phase diagram also shows the
regions of metastability (the purple, blue, yellow, and brown
regions), where there is more than one equilibrium solution to
Eqs.~(\ref{Exi}--\ref{Ekappa3}). We found the boundaries of these
regions by brute force; we systematically varied $\bar v_1$ and
$c_2$, solving the equilibrium equations for each pair of values
and noting when the number of solutions changed. The twist-stretch
coupling $k_5$ tilts the total region of metastability, whereas in
the simple model the region of metastability is vertical, $\bar
v_1^2<4/27$ (see Fig 2 of ~\cite{SrigirirajuPowers2005}). The 
phase diagram also has a dotted line showing where the two 
possible solutions of left-handed and right-handed filaments 
have equal energy.

The full model yields many possibilities for transitions,
including discontinuous transitions between straight states of
opposite handedness, discontinuous transitions between helical
states of opposite handedness, and continuous transitions between
straight states and helical states of the same handedness. These
transitions are present in the simple model. The full model also
predicts discontinuous transitions from a straight state to a
helical state of opposite handedness, which do not occur in the
simple model.

\begin{figure}
\includegraphics[height=2 in]{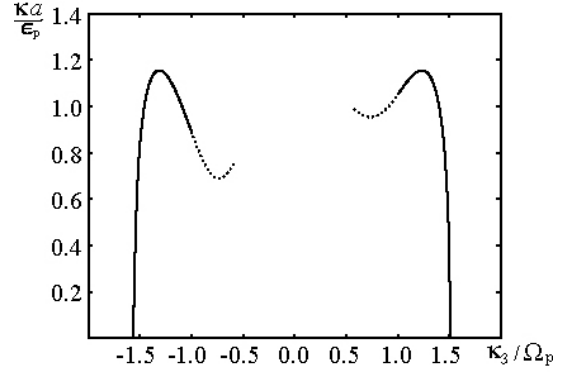}
\caption{Curvature vs. twist, in dimensionless units, for
$\gamma=1.8$, $\tilde u/v=1$, $\bar k_\mathrm{s}=8$, $\bar k_5=1$
and $\bar u_1=-0.5$. The parameters $\bar v_1$ and $c_2$ vary along the
dash-dot line of Fig.~\ref{pdfig}. The dotted lines corresponds 
to metastable states for which the lower energy states are of opposite handedness.}\label{cvtfig}
\end{figure}

As mentioned above, changes in solvent condition change the values
of the material parameters. To illustrate this effect, in
Fig.~\ref{cvtfig} we plot $\kappa$ vs. $\kappa_3$ for values of
$(\bar v_1,c_2)$ along the dashed line of Fig.~\ref{pdfig}. There
is a simultaneous jump in curvature and twist of the filament.
Note that $\kappa$ rises so steeply with twist that $\kappa$ may
appear to be undergoing an abrupt jump with small changes in
parameters. We also point out that the value of $\gamma$
determines the maximum pitch angle that can be attained by the
helical filament. Changing the value of
$\gamma=a\Omega_\mathrm{p}/\epsilon_\mathrm{p}$ amounts to
changing the relative positions of the minima in the nonlinear
potentials~(\ref{Usieqn}) and~(\ref{Ut}). If $\gamma$ is small,
there is a wider range of possible pitch angles. We chose
$\gamma=1.8$ for Figs.~\ref{pdfig} and~\ref{cvtfig}, which gives a
maximum pitch angle of about $35^\circ$. Higher pitch angle
polymorphs such as curly I and curly II require lower values of
$\gamma$.

\subsection{Response to moment and force}

\begin{figure}
\includegraphics[height=1.85in]{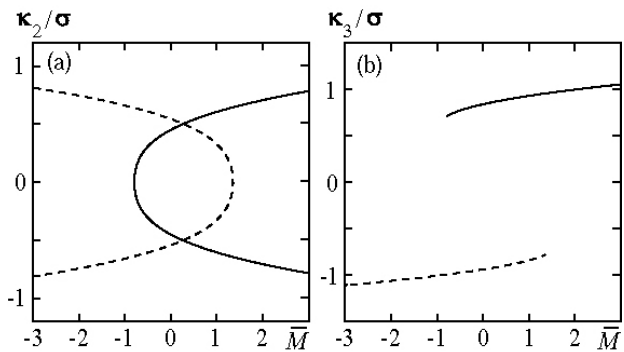}
\caption{(a) Signed curvature $\kappa_2$ vs. $\bar{M}$ for $\tilde
u=v$, $\bar k_\mathrm{s}=8$, $\bar k_5=1$, $c_2=4$, $\bar
v_1=-0.25$, and $\bar u_1=-0.5$. The dashed lines correspond to
left-handed states. (b) Twist $\kappa_3$ vs.
$\bar{M}$ with parameters as in (a).} \label{momentfullfig}
\end{figure}

We now consider the response of a filament to external moment and
force in the full model. Since the filament is extensible, there
is an additional contribution to the expression~(\ref{Fdr}) for
the work done by an external force:
\begin{equation}
\mathbf{F}\cdot\left[\delta\mathbf{r}(L) -
\delta\mathbf{r}(0)\right]=-F\int^L_0\left(\sin\theta
\delta\theta-\cos\theta\delta\epsilon\right)\mathrm{d}s,\label{Fdr2}
\end{equation}
where we have used the approximation
$\mathrm{d}S\approx\mathrm{d}s$. In addition to the Euler-Lagrange
equations derived for external moment in the simple model,
Eqs.~(\ref{el1}--\ref{bcphis}), suitably modified to include both
an external moment and force, there is a new equation
corresponding to the stretch degree of freedom $\epsilon$,
\begin{equation}
\frac{\partial V}{\partial\epsilon} -\frac{{\mathrm
d}}{\mathrm{d}s}\frac{\partial V}
{\partial \dot{\epsilon}}-F\cos\theta=0,\label{eq1}\\
\end{equation}
as well as natural boundary conditions
\begin{equation}
\left[\frac{\partial V}{\partial\dot\epsilon} \right]_{s=0}=\left[
\frac{\partial V}{\partial\dot\epsilon}
\right]_{s=L}=0.\label{bcxi_full}
\end{equation}
The argument leading to $\dot{\phi}=0$ in the simple model applies
here as well, leading to
\begin{eqnarray}
\frac{\epsilon^3}{\epsilon_\mathrm{p}^3}+c_1
\frac{\epsilon}{\epsilon_\mathrm{p}} -c_2-\frac{\bar
k_5\gamma\dot\psi\cos\theta}{\Omega_\mathrm{p}}+g_1&=
&\bar{F}\epsilon_\mathrm{p}\cos\theta\label{Exi1}\\
\frac{\dot\psi\sin\theta}{\Omega_\mathrm{p}} \left(\frac{\bar
k_5\gamma\epsilon}{\epsilon_\mathrm{p}} -g_2+g_3\cos\theta\right)
&=&-\bar{F}\sin\theta\label{Etheta}\\
g_2\cos\theta-\frac{\bar k_5\gamma\epsilon}{\epsilon_\mathrm{p}}
\cos\theta+g_3\sin^2\theta &=&\bar{M}\label{Epsi}
\end{eqnarray}
where  $ \bar{F}=F/(11u\epsilon_\mathrm{p}^4)$, $\bar{M}=
M\Omega_\mathrm{p}/(11u\epsilon_\mathrm{p}^4)$, and
\begin{eqnarray}
g_1&=&(3\gamma^2\epsilon\dot\psi^2\sin^2\theta)/
(2\epsilon_\mathrm{p}\Omega_\mathrm{p}^2)\label{g1}\\
g_2&=&\frac{v\gamma^4}{11u}\left[\frac{\dot\psi\cos\theta}{\Omega_\mathrm{p}}
\left(\frac{\dot\psi^2\cos^2\theta}{\Omega_\mathrm{p}^2}-1\right)
-\bar
v_1\right]\label{g2}\\
g_3&=&\frac{\gamma^2\dot\psi}{8\Omega_\mathrm{p}}
\left(-4+\frac{12\epsilon^2}{\epsilon_\mathrm{p}^2} +
\frac{3\gamma^2\dot\psi^2\sin^2\theta}{\Omega_\mathrm{p}^2}\right).\label{g3}
\end{eqnarray}
We use the numerical continuation method to solve these equations
for the cases of applied moment with $\bar F=0$, and applied force
with $\bar M=0$. We use the values of $\bar v_1$ and $c_2$
corresponding to the point X in Fig.~\ref{pdfig}, and the same
parameter values for $\gamma$, $\tilde u$, $v$, $\bar
k_\mathrm{s}$, $k_5$, and $\bar u_1$ used in
section~\ref{simpmod}. The point X corresponds to a stable
left-handed normal helix, and a metastable right-handed helix; we
study the response of the left-handed helix.
Figure~\ref{momentfullfig} shows that the left-handed helical
filament changes its shape smoothly with increasing moment, going
through helical metastable states until is straightens out, and
then abruptly changes to a right-handed filament. Note the 
qualitative similarity with the predictions of the simple model, 
Fig.~\ref{k2k3Ifig}. An important difference is that the asymmetry 
of the potential in the full model favors one handedness over the other; 
for example, at $M=0$, the right-handed solution of the full model 
has higher energy than the left-handed solution.

\begin{figure}
\includegraphics[height=1.8in]{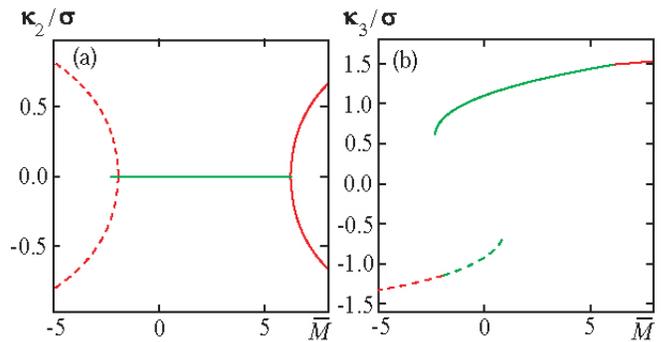}
\caption{(a) Signed curvature $\kappa_2$ vs. $\bar{M}$ for $\tilde
u=v$, $\bar k_\mathrm{s}=8$, $\bar k_5=1$, $c_2=15$, $\bar
v_1=-0.25$, and $\bar u_1=-0.5$. The red and green curves indicate the 
helical and straight filaments, respectively. The dashed lines correspond to
left-handed states. (b) Twist $\kappa_3$ vs. $\bar{M}$ with
parameters as in (a).} \label{momentstgfullfig}
\end{figure}

The response to moment of a filament which is straight when $M=0$
is shown in Fig.~\ref{momentstgfullfig}. The parameters $\bar v_1$
and $c_2$ are chosen to correspond to the point Y in
Fig.~\ref{pdfig}, which represents a stable right-handed straight
filament. At these parameter values there is also a metastable
left-handed straight filament. As the moment increases from zero,
the twist increases until eventually the straight filament
destabilizes at a critical torque, and there is a continuous
transition to a right-handed helix. When the moment decreases from
zero, the straight filament untwists and and a left-handed helix
forms at a critical torque. Due to the asymmetry of the potential
for these parameter values, the interval of moments for which
left-handed straight filaments exist is a subset of the interval
of moments for which right-handed straight filaments exist.
Nevertheless, Fig.~\ref{momentstgfullfig} is qualitatively similar
to Fig.~\ref{k2k3stfig}.

\begin{figure}
\includegraphics[height=1.8in]{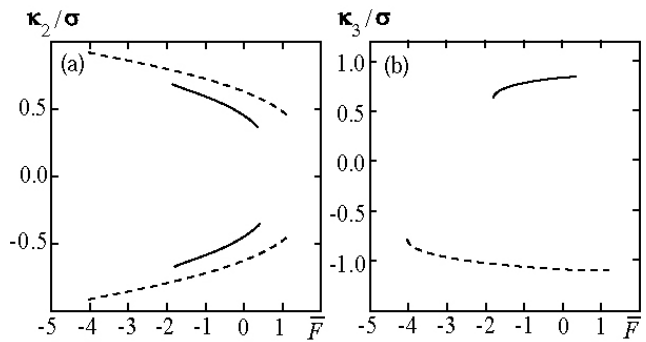}
\caption{(a) Signed curvature $\kappa_2$ vs. $\bar{F}$ for $\tilde
u=v$, $\bar k_\mathrm{s}=8$, $\bar k_5=1$, $c_2=15$, $\bar
v_1=-0.25$, and $\bar u_1=-0.5$. The dashed lines correspond to
left-handed states. (b) Twist $\kappa_3$ vs.
$\bar{F}$ with parameters as in (a).} \label{forcefullfig}
\end{figure}

Finally, Fig.~\ref{forcefullfig} shows the response of the
left-handed filament at X to an applied force, showing the same
qualitative features we found in the simple model,
Fig.~\ref{forcesimplfig}.

\section{Discussion and conclusion}\label{discusssec}

We have introduced a new coarse-grained model for polymorphic
transitions in bacterial flagella. The key elements of the model
are a double-well potential for protofilament stretch, a
double-well potential for twist arising from lateral interactions
between neighboring protofilaments, elastic mismatch between the
protofilaments and the inner core, and cooperative interactions
between subunits on the same protofilament.  Motivated by the
chiral structure of flagellar filaments, we have also included a
potential energy for twist-stretch coupling in the full version of
our model. This potential energy is in marked contrast with the
tight coupling between filament twist and protofilament stretch
assumed in~\cite{calladine1975} and~\cite{hasegawa_etal1998},
which leads to a discrete set of possible values for the curvature
and twist of a filament in the absence of loading. Our softer
coupling in the full model ($k_5\ne0$) and complete decoupling
($k_5=0$) in the simple model lead to a continuous variation of
curvature and twist as a function of the material parameters, with
discontinuous transitions only at special critical values. Our
prediction calls for new single-molecule experiments in which the
shape of a \textit{single} filament is observed as solvent
conditions are carefully varied.

The main result of our work is the calculation of the response of 
a polymorphic filament to moments and forces applied at the ends. In the
simplified version of our theory, the assumption of a stiff
central core leads to semi-analytic formulas for the shape as a
function of moment. These formulas allowed us to reveal a host of
possible distinct scenarios, such as the moment-driven continuous
unwinding of a helical filament to a straight shape, followed by
an abrupt transition to a helix of opposite handedness as the
moment increases (Fig.~\ref{k2k3Ifig}, blue curves). 
Another important generic prediction 
of our model is that the curvature
and the twist both jump at the same critical value of moment.
Although our model makes clear predictions for the qualitative
behavior of filaments, a critical requirement for testing the
quantitative validity of our theory is the determination of the
values of the material parameters such as $u$, $v$, $\bar v_1$,
and $\epsilon_0$. These may be measured by bending and twisting
straight filaments, or perhaps may be computing using the
techniques developed in~\cite{flynn_ma2004}. We can make a rough 
order-of-magnitude estimate of the moment required to cause a 
polymorphic transformation by observing that the critical moment 
in the simple model is approximately $M\approx  u a^4 \kappa_0^2$, 
where we have assumed for simplicity that $ u\approx v$, 
$\kappa_0\approx\Omega_\mathrm{p}$,  $\sigma\approx\kappa_0^2$, 
and we have also dropped overall numerical factors 
(such as the distinction between $u$ and $\tilde u$).  
Since the bending stiffness $A\approx u a^4\kappa_0^2$, 
we may write $M\approx A\kappa_0$ for the characteristic moment. 
Using $\kappa_0\approx 1$\,$\mu$m, $A\approx 10^{-24}\,$Nm$^2$ 
(see~\cite{kim_powers2005} and references therein) yields a 
characteristic torque for polymorphic transition of $10^{-18}$\,Nm. 
This torque is comparable to the hydrodynamic torque experienced by 
a rotating flagellum, which we can estimate by computing the the 
torque on the cell body. Assuming the cell body rotates at $\omega=10$\,Hz~\cite{macnab1977} 
and is a sphere of radius $R=1$\,$\mu$m leads to a hydrodynamic torque 
of the order of $M=8\pi\eta R^3\omega\approx10^{-18}$\,Nm~\cite{LandauFM}. 
These values match well with those of Hotani, who estimated the 
orders of magnitude for the transformation of a normal to semi-coiled 
filament to be $10^{-18}$\,Nm, and $10^{-19}$\,Nm for the reverse 
transformation~\cite{hotani1982}.
 
The shape-moment curves predicted by our theory may be readily
tested by new micromanipulation experiments in which a filament is
subject to prescribed moments or forces. It would also be
interesting to vary the solvent conditions while doing these
measurements. Also, since our theory makes predictions about the
sign of $\kappa_2$, it would be useful to experimentally track the
sign of $\kappa_2$ during a transition. For example, for a
transition such as is depicted in Fig.~\ref{k2k3IIfig}(a), a bead
which is stuck to a protofilament which is on the inside of the
helix before the transition would end up on the outside
afterwards.

Our theory also suggests new directions for theoretical work. In
this article we have considered boundary conditions of fixed
moment or force. If instead we fixed the positions and
orientations of the ends of the filament (``hard boundary
conditions"), then the filament could take on a shape in which two
different polymorphic states coexist (see Fig. 1(a)
of~\cite{goldstein_etal2000} for a similar experimental example
with \textit{free} filaments with subunits of mixed type). We
could use our theory to analyze these states, generalizing the
calculations for planar beams with a bistable potential for
curvature carried out in~\cite{james1981}. Another important
generalization of our theory would be to study the dynamics of
polymorphic transitions, which has been studied in the context of
a different continuum model in~\cite{goldstein_etal2000}
and~\cite{CoombsHuberKesslerGoldstein2002}. Finally, it is natural
to attempt to go beyond our protofilament-based picture and treat
every subunit on equal footing, and try to adapt ideas developed
for understanding martensitic transformations to polymorphic
flagella~\cite{bhattacharya2004}.

\begin{acknowledgments}
We thank H. Berg, C. Calladine, D. DeRosier, A. Goriely, K. Namba,
A. Needleman, L. Turner, and C. Wolgemuth for helpful
conversations. We are especially grateful to Greg Huber for many
important early discussions. We thank C. Wolgemuth for providing
the code used to generate Fig.~\ref{circlehelix} and other
figures. This work is supported in part by National Science
Foundation grants CMS-0093658 and NIRT-0404031, and the Brown
University MRSEC. TRP thanks the Aspen Center for Physics for
hospitality while some of this work was completed.
\end{acknowledgments}

\appendix
\section{Euler angles}\label{euleranglessection}

The sequence of rotations defining the Euler angles is described
in section~\ref{momsec}. Carrying out these rotations on the
initial frame $\{\hat\mathbf{x},\hat\mathbf{y},\hat\mathbf{z}\}$
leads to
\begin{eqnarray}
\hat\mathbf{e}_1'&=&\cos\psi\hat\mathbf{x}+\sin\psi\hat\mathbf{y}\label{e1p}\\
\hat\mathbf{e}_2'&=&-\sin\psi\hat\mathbf{x}+\cos\psi\hat\mathbf{y}\label{e2p}\\
\hat\mathbf{e}_3'&=&\hat\mathbf{z}\label{e3p},
\end{eqnarray}
then
\begin{eqnarray}
\hat\mathbf{e}_1''&=&\cos\psi\cos\theta\hat\mathbf{x}+\sin\psi\cos\theta\hat\mathbf{y}
-\sin\theta\hat\mathbf{z}\label{e1pp}\\
\hat\mathbf{e}_2''&=&-\sin\psi\hat\mathbf{x}+\cos\psi\hat\mathbf{y}\label{e2pp}\\
\hat\mathbf{e}_3''&=&\cos\psi\sin\theta\hat\mathbf{x}+\sin\psi\sin\theta\hat\mathbf{y}
+\cos\theta\hat\mathbf{z}\label{e3pp},
\end{eqnarray}
and finally
\begin{eqnarray}
\hat\mathbf{e}_1&=&(-\sin\psi\sin\phi+\cos\psi\cos\phi\cos\theta)
\hat\mathbf{x} \nonumber\\
&+&(\cos\psi\sin\phi + \sin\psi\cos\phi\cos\theta) \hat\mathbf{y}
\nonumber\\ &-&\cos\phi\sin\theta\hat \mathbf{z}\label{e1euler}\\
\hat\mathbf{e}_2&=&-(\sin\psi\cos\phi+\cos\psi\sin\phi\cos\theta)
\hat\mathbf{x} \nonumber\\
&+&(\cos\psi\cos\phi - \sin\psi\sin\phi\cos\theta) \hat\mathbf{y}
\nonumber\\ &+&\sin\phi\sin\theta\hat \mathbf{z}\label{e2euler}\\
\hat\mathbf{e}_3&=&\cos\psi\sin\theta \hat\mathbf{x}
+\sin\psi\sin\theta \hat\mathbf{y}
+\cos\theta\hat\mathbf{z}.\label{e3euler}
\end{eqnarray}

Plugging Eqs.~(\ref{e1euler}--\ref{e3euler}) into
Eq.~(\ref{onframeeqn}),
$\mathrm{d}\hat\mathbf{e}_\mu/\mathrm{d}s=\bm{\kappa}
\times\hat\mathbf{e}_\mu$, yields Eqs.~(\ref{k1}--\ref{k3}), which
show that the $\kappa_\mu$ depend on $\theta$, $\phi$, $\dot\psi$,
and $\dot\phi$. Furthermore,
\begin{equation}
\kappa^2=\dot\theta^2+\sin^2\theta\dot\psi^2\label{kappasquared}
\end{equation}
and
\begin{eqnarray}
\dot\kappa_1^2+\dot\kappa_2^2&=& \sin^2\theta\dot\phi^2\dot\psi^2
+\dot\theta^2{\left(\dot\phi  -
       \cos\theta \dot\psi \right)}^2\nonumber\\
       &+&
  2\sin\theta\dot\phi\dot\psi\ddot\theta +
  \ddot\theta^2 \nonumber\\&-& 2\sin\theta\dot\theta
   \left( \dot\phi - \cos\theta\dot\psi\right)
   \ddot\psi
    + \sin^2\theta\ddot\psi^2
   \label{coopfull}
\end{eqnarray}
Therefore, $\kappa^2$ and $\dot\kappa_1^2+\dot\kappa_2^2$ are both
independent of $\psi$, $\phi$, and $\ddot\phi$. Since $\kappa_3$
is also independent of these variables [see Eq.~(\ref{k3})], the
energy density $U$ is also.


\begin{thebibliography}{10}

\bibitem{eaton_etal1999}
W.A. Eaton, E.R. Henry, J.~Hofrichter, and A.~Mozzarelli,
Nature Struct. Biol. {\bf 6}, 351 (1999).

\bibitem{lawsonsigler1988}
C.L. Lawson and P.B. Sigler, 
Nature {\bf 333}, 869 1988.

\bibitem{krausevolzlipscomb1985}
K.L. Krause, K.W. Volz, and W.N. Lipscomb,
Proc. Natl. Acad. Sci. USA {\bf 82}, 1643 (1985).

\bibitem{brayduke2004}
D.~Bray and T.~Duke,
Annu. Rev. Biophys. Biomol. Struct. {\bf 33}, 53 (2004).

\bibitem{macnab1996}
R.M. Macnab,
In {\em \textit{Eschericia
  coli} and \textit{Salmonella}}, edited by F.C.~Neidhardt \textit{et al.}, (American Society for Microbiology,
  Washington, D.C., 1996), pp. 123.

\bibitem{asakura1970}
S.~Asakura,
Advan. in Biophys. {\bf 1}, 99 (1970).

\bibitem{turner_ryu_berg2000}
L.~Turner, W.S. Ryu, and H.C. Berg,
J. Bacteriol. {\bf182}, 2793 (2000).

\bibitem{macnab_ornston1977}
R.M. Macnab and M.K. Ornston,
J. Mol. Biol. {\bf 112}, 1 (1977).

\bibitem{hotani1982}
H.~Hotani,
J. Mol. Biol. {\bf 156}, 791 (1982).

\bibitem{berg2004}
H.C. Berg,
\newblock {\em \textit{E. coli} in Motion}
(Springer-Verlag, New York, 2004).

\bibitem{kim_etal2003}
M.J. Kim, J.C. Bird, A.J. {Van Parys}, K.S. Breuer, and T.R. Powers,
Proc. Natl. Acad. Sci. USA {\bf 100}, 15481 (2003).

\bibitem{kamiyaasakura1976}
R.~Kamiya and S.~Asakura,
J. Mol. Biol. {\bf 106}, 167 (1976).

\bibitem{hasegawa_etal1982}
K.~Hasegawa, R.~Kamiya, and S.~Asakura,
J. Mol. Biol. {\bf 160}, 609 (1982).

\bibitem{SrigirirajuPowers2005}
S.V. Srigiriraju and T.R. Powers,
Phys. Rev. Lett. {\bf 94}, 248101 (2005).

\bibitem{yamashita_etal1998}
I.~Yamashita, K.~Hasegawa, H.~Suzuki, F.~Vonderviszt, Y.~Mimori-Kiyosue, and
  K.~Namba,
Nature Struct. Biol. {\bf 5}, 125 (1998).

\bibitem{crick_watson1956}
F.H.C. Crick and J.D. Watson,
Nature {\bf 177}, 473 (1956).

\bibitem{fraenkel-conrat_williams1955}
H.~Fraenkel-Conrat and R.C. Williams,
Proc. Natl. Acad. Sci. USA {\bf 41}, 690 (1955).

\bibitem{hyman_trachtenberg1991}
H.C. Hyman and S.~Trachtenberg,
J. Mol. Biol. {\bf 220}, 79 (1991).

\bibitem{mimori_etal1995}
Y.~Mimori, I.~Yamashita, K.~Murata, Y.~Fujiyoshi, K.~Yonekura, C.~Toyoshima,
  and K.~Namba,
J. Mol. Biol. {\bf249}, 69 (1995).

\bibitem{yoshioka_aizawa_yamaguchi1995}
K.~Yoshioka, S.~Aizawa, and S.~Yamaguchi,
J. Bacteriol. {\bf 177}, 1090 (1995).

\bibitem{kanto_etal1991}
S.~Kanto, H.~Okino, S.I. Aizawa, and S.~Yamaguchi,
J. Mol. Biol. {\bf 219}, 471 (1991).

\bibitem{mimori-kiyosue_etal1996}
Y.~Mimori-Kiyosue, F.~Vonderviszt, I.~Yamashita, Y.~Fujiyoshi, and K.~Namba,
Proc. Natl. Acad. Sci. USA {\bf93}, 15108 (1996).

\bibitem{vonderviszt_aizawa_namba1991}
F.~Vonderviszt, S.I. Aizawa, and K.~Namba,
J. Mol. Biol. {\bf 221}, 1461 (1991).

\bibitem{samatey_etal2001}
F.A. Samatey, K.~Imada, S.~Nagashima, F.~Vonderviszt, T.~Kumasaka, M.~Yamamoto,
  and K.~Namba,
Nature {\bf 410}, 331 (2001).

\bibitem{yonekura_etal2003}
K.~Yonekura, S.~Maki-Yonekura, and K.~Namba,
Nature {\bf 424}, 643 (2003).

\bibitem{calladine1975}
C.R. Calladine,
Nature {\bf 255}, 121 (1975).

\bibitem{calladine1976}
C.R. Calladine,
J. Theoret. Biol. {\bf 57}, 469 (1976).

\bibitem{calladine1978}
C.R. Calladine,
J. Mol. Biol. {\bf 118}, 457 (1978).

\bibitem{hasegawa_etal1998}
K.~Hasegawa, I.~Yamashita, and K.~Namba,
Biophys. J. {\bf 74}, 569 (1998).

\bibitem{goldstein_etal2000}
R.E. Goldstein, A.~Goriely, G.~Huber, and C.~W. Wolgemuth,
Phys. Rev. Lett. {\bf 84}, 1631 (2000).

\bibitem{CoombsHuberKesslerGoldstein2002}
D.~Coombs, G.~Huber, J.O. Kessler, and R.E. Goldstein,
Phys. Rev. Lett. {\bf 89}, 118102 (2002).

\bibitem{landau_lifshitz_elas}
L.D. Landau and E.M. Lifshitz,
\newblock {\em Theory of Elasticity}
(Pergamon Press, Oxford, 1986).

\bibitem{marko_siggia1994}
J.F. Marko and E.D. Siggia, 
Macromolecules {\bf 27}, 981 (1994).

\bibitem{namba_vonderviszt1997}
K.~Namba and F.~Vonderviszt,
Q. Rev. Biophys. {\bf30}, 1 (1997).

\bibitem{kamien2002}
R.D. Kamien,
Rev. Mod. Phys. {\bf 74}, 953 (2002).

\bibitem{love1944}
A.E.H. Love,
\newblock {\em A treatise on the mathematical theory of elasticity}
(Dover Publications, New York, 4th edition, 1944).

\bibitem{keller1977}
H.~B. Keller,
\newblock {\em Numerical solution of bifurcation and nonlinear eigenvalue
  problems, in Applications of bifurcation theory}, edited by P. H.
  Rabinowitz,
(Academic Press, New York, 1977).

\bibitem{flynn_ma2004}
T.C. Flynn and J.P. Ma,
Biophys. J. {\bf 86}, 3204 (2004).

\bibitem{kim_powers2005}
M.J. Kim and T.R. Powers,
Phys. Rev. E {\bf 71}, 021914 (2005).

\bibitem{macnab1977}
R.M. Macnab,
Proc. Natl. Acad. Sci. USA {\bf 74}, 221 (1977).

\bibitem{LandauFM}
L.D. Landau and E.M. Lifshitz,
\newblock {\em Fluid Mechanics}
(Butterworth-Heinemann Ltd., Oxford, 2nd edition, 1987).

\bibitem{james1981}
R.D. James,
J. Elasticity {\bf 11}, 239 (1981).

\bibitem{bhattacharya2004}
K.~Bhattacharya,
\newblock {\em Microstructure of Martensite}
(Oxford University Press, Oxford, 2004).

\end{thebibliography}

\end{document}